\documentclass[journal]{vgtc}                     

\onlineid{1281}

\vgtccategory{Analytics \& Decisions}

\title{True (VIS) Lies: Analyzing How Generative AI Recognizes Intentionality, Rhetoric, and Misleadingness in Visualization Lies}

\author{%
  \authororcid{Graziano Blasilli}{0000-0003-3339-6403} and
  \authororcid{Marco Angelini}{0000-0001-9051-6972}
}

\authorfooter{
  \item
  	Graziano Blasilli is with Sapienza University of Rome, Rome, Italy. 
  \item
  	Marco Angelini is with Link Campus University, Rome, Italy.
}

\abstract{%
This study investigates the ability of multimodal Large Language Models (LLMs) to identify and interpret misleading visualizations, and recognize these observations along with their underlying causes and potential intentionality. Our analysis leverages concepts from visualization rhetoric and a newly developed taxonomy of authorial intents as explanatory lenses. 
We formulated three research questions and addressed them experimentally using a dataset of 2,336 COVID-19-related tweets, half of which contain misleading visualizations, and supplemented it with real-world examples of perceptual, cognitive, and conceptual errors drawn from VisLies, the IEEE VIS community event dedicated to showcasing deceptive and misleading visualizations.
To ensure broad coverage of the current LLM landscape, we evaluated 16 state-of-the-art models. Among them, 15 are open-weight models, spanning a wide range of model sizes, architectural families, and reasoning capabilities. 
The selection comprises small models, namely Nemotron-Nano-V2-VL (12B parameters), Mistral-Small-3.2 (24B), DeepSeek-VL2 (27B), Gemma3 (27B), and GTA1 (32B); medium-sized models, namely Qianfan-VL (70B), Molmo (72B), GLM-4.5V (108B), LLaVA-NeXT (110B), and Pixtral-Large (124B); and large models, namely Qwen3-VL (235B), InternVL3.5 (241B), Step3 (321B), Llama-4-Maverick (400B), and Kimi-K2.5 (1000B). In addition, we employed OpenAI GPT-5.4, a frontier proprietary model.
To establish a human perspective on these tasks, we also conducted a user study with visualization experts to assess how people perceive rhetorical techniques and the authorial intentions behind the same misleading visualizations. 
This allows comparison between model and expert behavior, revealing similarities and differences that provide insights into where LLMs align with human judgment and where they diverge. Results show that while LLMs have a basic ability to identify misleading elements, their performance is influenced by the prompting strategy. Additionally, the findings indicate that LLMs can also reason, to varying extents, about the rhetorical aspects and intentionality of visual misinformation.

}


\teaser{
    \centering
    \vspace{2mm}
    \includegraphics[width=\linewidth]{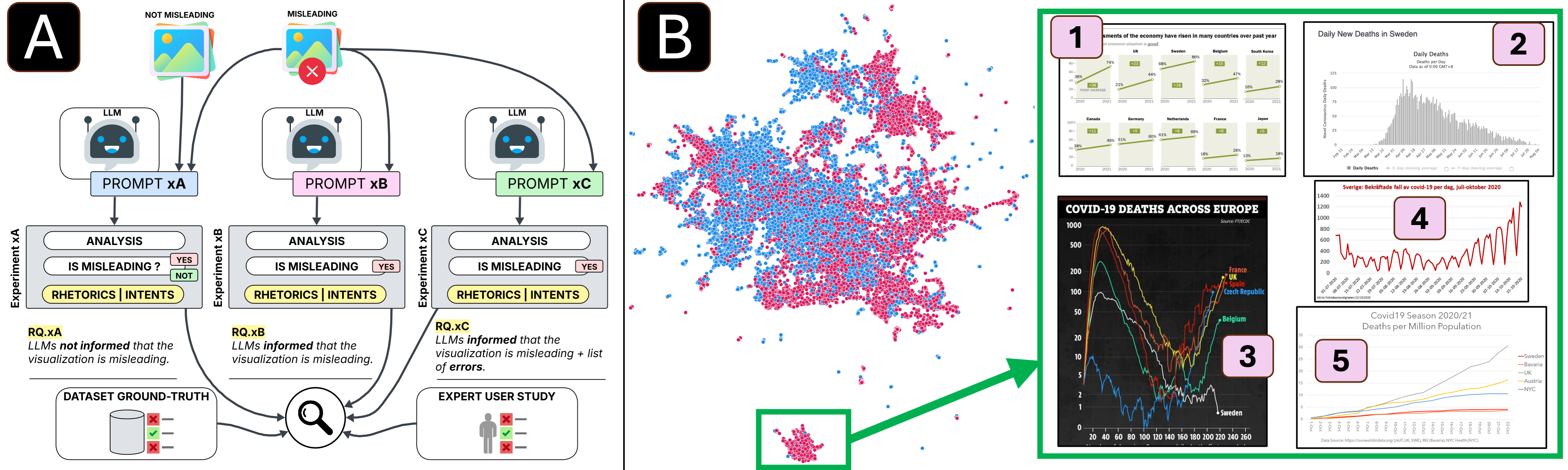}
    \caption{
(A) Overview of the study design. Three conditions govern the prior knowledge given to the models: in condition A, the model is not informed about the visualization's misleadingness; in condition B, it is told that the visualization is misleading; in condition C, it additionally receives the ground-truth error list. 
(B) UMAP projection of model explanations for why a visualization is perceived as misleading (experiment \experiment{1A}). Each point represents one tweet-model pair; color encodes ground truth (blue: non-misleading, red: misleading). The projection forms a single densely packed cloud, confirming that models produce semantically similar explanations regardless of ground truth. The green box highlights a cluster of five tweets that share a narrow discourse pattern: cross-country COVID-19 trend charts are used to argue about lockdown policy, often featuring Sweden, with argumentative captions that frame the intended conclusion before the reader inspects the data. 
The cluster is generated by topic and discovered errors, and most models attribute its misleadingness to Information Access rhetoric, where cherry-picking of countries, time windows, and comparison baselines appear to be the main drivers.
The highlighted tweets are: 1) \tlink{1421622713509048323}, 2) \tlink{1291399556224221186}, 3) \tlink{1322461003209166849}, 4) \tlink{1319989877803503617}, and 5) \tlink{1319519135685054466}.
}
\label{fig:teaser}
}

\renewcommand{\manuscriptnotetxt}{}

\graphicspath{{figs/}{figures/}{pictures/}{images/}{./}} 

\usepackage{tabu}                      
\usepackage{booktabs}                  
\usepackage{lipsum}                    
\usepackage{mwe}                       
\usepackage{ccicons}                   

\usepackage{mathptmx}                  

\usepackage{tcolorbox}
\usepackage{array}
\usepackage{booktabs, multirow}
\usepackage{tabularx}
\usepackage[table,dvipsnames]{xcolor}
\definecolor{darkgreenexcel}{HTML}{38761D}
\definecolor{lightgreenexcel}{HTML}{279B3A}

\usepackage{subcaption}
\usepackage{graphicx} 
\usepackage{lipsum}
\usepackage{fontawesome6}
\usepackage{fvextra} 
\usepackage{tcolorbox}
\tcbuselibrary{skins,breakable}
\usepackage{amsmath}

\newcommand{\para}[1]{\vspace{2mm}\noindent\textbf{#1}\hspace{0.5em}} 

\DeclareRobustCommand{\rquestion}[1]{\protect{%
\renewcommand*{\sfdefault}{PTSansNarrow-TLF}%
\sffamily\bfseries
\fontsize{8.4pt}{11pt}\selectfont
RQ.#1%
}}

\DeclareRobustCommand{\RQuestion}[1]{\protect{%
\renewcommand*{\sfdefault}{PTSansNarrow-TLF}%
\sffamily\bfseries
RQ.#1%
}}

\DeclareRobustCommand{\rcond}[1]{\protect{%
\renewcommand*{\sfdefault}{PTSansNarrow-TLF}%
\sffamily\bfseries #1%
}}

\DeclareRobustCommand{\experiment}[1]{\protect{%
\renewcommand*{\sfdefault}{PTSansNarrow-TLF}%
\sffamily\bfseries E.#1%
}}

\DeclareRobustCommand{\prompt}[1]{\protect{%
\renewcommand*{\sfdefault}{PTSansNarrow-TLF}%
\sffamily\bfseries P.#1%
}}

\newcommand{\model}[1]{\protect{%
\texttt{#1}%
}}

\newcommand{\deepseek}{\model{deepseek}}
\newcommand{\gemma}{\model{gemma}}
\newcommand{\glm}{\model{glm}}
\newcommand{\gta}{\model{gta}}
\newcommand{\intern}{\model{intern}}
\newcommand{\kimi}{\model{kimi}}
\newcommand{\llava}{\model{llava}}
\newcommand{\maverick}{\model{maverick}}
\newcommand{\mistral}{\model{mistral}}
\newcommand{\molmo}{\model{molmo}}
\newcommand{\nemotron}{\model{nemotron}}
\newcommand{\pixtral}{\model{pixtral}}
\newcommand{\qianfan}{\model{qianfan}}
\newcommand{\qwen}{\model{qwen}}
\newcommand{\stepthree}{\model{step3}}
\newcommand{\gpt}{\model{gpt}}

\newcommand{\hflink}[1]{\tiny{\href{https://huggingface.co/#1}{~\faLink}}}

\newcommand{\tlink}[1]{\protect{%
\small\href{https://x.com/i/web/status/#1}{\texttt{#1}}%
}}

\newcommand{\textbfi}[1]{{\textit{\bfseries#1}}}

\definecolor{ibmblue}{HTML}{648FFF}
\definecolor{ibmpurple}{HTML}{785EF0}
\definecolor{ibmpink}{HTML}{DC267F}
\definecolor{ibmorange}{HTML}{FE6100}
\definecolor{ibmyellow}{HTML}{FFB000}

\definecolor{colorA}{HTML}{C1E5F5} \definecolor{colorADark}{HTML}{156082}
\definecolor{colorB}{HTML}{F2CFEE} \definecolor{colorBDark}{HTML}{A02B93}
\definecolor{colorC}{HTML}{D9F2D0} \definecolor{colorCDark}{HTML}{4EA72E}

\newtcbox{\cb}[1][]{%
  on line,
  nobeforeafter,      
  tcbox raise base,   
  arc=2pt,
  colback=white,
  colframe=blue,
  boxsep=0pt,
  left=1pt,right=1pt,top=0.3pt,bottom=0.3pt,
  #1
}

\newcommand{\cbA}{%
\cb[arc=1pt,colframe=colorA,colback=colorA]{\bfseries A}%
}

\newcommand{\cbB}{%
\cb[arc=1pt,colframe=colorB,colback=colorB]{\bfseries B}%
}

\newcommand{\cbC}{%
\cb[arc=1pt,colframe=colorC,colback=colorC]{\bfseries C}%
}

\definecolor{modelSmall}{HTML}{FEE726}
\definecolor{modelMedium}{HTML}{27AB82}
\definecolor{modelLarge}{HTML}{3C4F8A}
\usepackage[subtle]{savetrees}

\begin{document}


\maketitle

\section{Introduction}
\label{sec:introduction}

Misleading visualizations are a persistent source of misinformation, particularly on social media \cite{Lisnic2023,Lee2021,LisnicLex2024}. 
Their deceptive effect can arise from explicit design violations, such as truncated axes or inappropriate encodings, from reasoning errors, such as cherry-picking or causal inference, or from a combination of both \cite{LoGupta2022,Szafir2018}. 
As multimodal Large Language Models (LLMs) become increasingly embedded in everyday workflows, lay users are turning to them for help interpreting visualizations they encounter online \cite{10.1145/3711061}. This creates a concrete risk: if an LLM fails to recognize that a visualization is misleading, or mischaracterizes the nature of its deception, the user may trust a flawed chart for the wrong reasons.
Recent work has begun to benchmark LLMs for misleading visualization detection by assessing recognition accuracy and the effects of prompting strategies \cite{https://doi.org/10.1111/cgf.70137, Alexander2024, Lo2025}.
However, these contributions share a common focus on \textit{whether} a model can detect that something is wrong. They do not ask \textit{how} the deception is constructed, through which communicative strategies (visualization rhetoric \cite{Hullman2011}), or \textit{why} it may have been produced, whether through deliberate manipulation or unintentional factors such as limited visualization literacy.

This paper contributes to answering these deeper research questions. We evaluate 16 state-of-the-art multimodal LLMs, 15 open-weight and one proprietary, on their ability to (a) spontaneously detect misleading visualizations, (b) recognize the rhetorical techniques that shape their persuasive effect, and (c) attribute the authorial intents behind the deception. 
To structure the analysis, we contribute a novel taxonomy of nine authorial intents, grouped into intentional and unintentional categories. 
The evaluation is conducted on 2,336 COVID-19 tweets \cite{Lisnic2023} (half misleading) and supplemented with 130 examples from the VisLies gallery \cite{vislies}, across six experiments that progressively increase the prior knowledge provided to the models. A user study with 11 visualization experts provides a human reference for the rhetoric and intent attribution tasks.
The datasets, all the generated data from the tested LLMs, and a web-based explorer of results are available at {\small\url{https://github.com/XAIber-lab/truevislies}}.

\section{Background and Related Work}
\label{sec:related}

Large Language Models and Generative AI capabilities in supporting visualization have been extensively tested in the last three years~\cite{YE202443}. Several efforts have been identified for visualization generation~\cite{10541799,10670418,10.1145/3654992,10670425,10443572}, with a focus on generating correct visualizations that adhere to visualization literacy rules~\cite{10857634,https://doi.org/10.1111/cgf.70137,11261841}. While these works tested the LLMs' capability to avoid producing erroneous visualizations, their focus was not on detecting them or on the eventual misinterpretations they provide to a lay user.
Another part of the literature focused on supporting a user in the automatic or semi-automatic analysis and interpretation of charts~\cite{huang-etal-2024-lvlms,10988687}. Podo et al.~\cite{podo2024vrecslowcostllm4visrecommender} created a dedicated model, distilled from ChatGPT-4, to support the generation, explanation, and analysis of suggestions for lay users or domain experts coping with visual data representations. Islam et al.~\cite{DBLP:conf/emnlp/IslamRMLNH24} extensively tested LLMs and vision models in chart understanding and reasoning, highlighting a non-negligible loss in accuracy due to hallucinations, factual errors, and data bias. Dong and Crisan~\cite{Dong2025} took a step further, analyzing how the reasoning of VLMs and humans can align or diverge in their interpretation of chart characteristics (e.g., marks, channels).
Mukherjee et al. contributed ENCQA~\cite{11262793}, a benchmark of nine VLMs on 2,076 synthetic question-answer pairs, for six visual encoding channels and eight tasks. This work identifies potential shortcomings of VLMs for chart interpretation that may result in misleading answers, such as the struggle with encodings that require interpreting legends in relation to those that require reading values off an axis or making visual estimates of correlation.

A few works directly addressed the problem of potential misinterpretations of generated charts and the detection of their possible causes.
Pandey and Ottley~\cite{https://doi.org/10.1111/cgf.70137} benchmarked multiple AI models, testing their capabilities to interpret charts and eventually discovering low performance in spotting misleading visualizations, with just 30\% accuracy. Alexander et al.~\cite{Alexander2024} show that GPT-4 models can detect misleading visualizations~\cite{LoGupta2022,Szafir2018} with varying accuracy depending on the used prompt-engineering technique. The authors share our goal, but they limited testing to GPT-4, whereas we expand on this work by considering additional models. Moreover, we take a step forward by attempting to detect the potentially malicious nature of the misleading visualization and the visual rhetoric strategy employed. Lo and Qu~\cite{Lo2025} also delve into this problem, focusing on defining nine types of prompts, from simple to complex, to support the detection of misleading visualizations. 
Das and Mueller~\cite{11269882} propose a visual analytic solution, MisVisFix. It leverages LLMs (Claude and ChatGPT) to support human experts and fact-checkers in detecting, explaining, and correcting misleading visualizations.
Mahbub et al.~\cite{Mahbub2025} very recently assessed the influence that misleading visualizations have on VLMs through a Wilcoxon signed-rank test comparing original and misleading chart pairs on eight visual coordinates, leading to misleading behavior.
Finally, Tonglet et al.\cite{Tonglet2025} propose two inference-time methods, table-based QA and redrawing the visualization, to improve the accuracy of VLMs up to $\approx 20 \%$ in detecting misleading visualizations.

While the authors of these contributions focused on detecting or improving detection capabilities, eventually linking them to potential symptoms mostly linked to visual mappings, in our work, we want to test more the case of a lay user potentially exposed to misinformation (looking more at the simple prompts than more complex ones) and the identification of potential root causes and rhetoric used to convey the erroneous information. Additionally, Szafir~\cite{Szafir2018} demonstrates that, in most cases, misleading visualizations are not primarily due to violations of visualization design guidelines, but rather to other factors. Among them, we identified the use of visualization rhetoric, which is not dissimilar to what can happen in natural language, where classic rhetorical constructs are used to deliver a message~\cite{raffini2025persuasive}.


Understanding misleading visualizations requires distinguishing between two complementary perspectives that together characterize how and why a visualization fails to represent and report data accurately.
First, we briefly discuss established taxonomies of \textit{misleading errors}, which classify the specific design violations and reasoning errors that can make a visualization inaccurate or deceptive.
Second, we summarize the techniques of \textit{visualization rhetoric}, which describes how choices in data selection, visual encoding, and linguistic framing shape the interpretation of the visualizations, often in ways that go beyond explicit errors.
Together, these two perspectives inform the design of our study.
The third conceptual pillar of our work is a novel taxonomy of \textit{authorial intents} behind misleading visualizations, presented in \cref{sec:intents}.


\subsection{Misleading Errors}
\label{sec:errors_def}
Lo et al. \cite{LoGupta2022} introduced a taxonomy of 74 issues that generate misleading visualizations. Building on this foundation, Lisnic et al. \cite{Lisnic2023} extended the scope of analysis by examining how charts mislead people in practice. They introduced a distinction between visual design violations and reasoning errors, highlighting that misleading effects often arise not only from how data is inadequately encoded visually (e.g., truncated axes, unclear encodings) but also from reasoning errors underlying the visualization (e.g., cherry-picking, causal inference). Their taxonomy includes 14 errors: seven \textit{visualization design violations} and seven \textit{reasoning errors}.
As supplemental material, we report the exact definitions of all error types we used in the prompts of our experiments, sourced from Lisnic et al.~\cite{Lisnic2023}.



\subsection{Visualization Rhetoric}
\label{sec:rhetoric_def}

When creating data visualizations, choices in data selection, visual encoding, language, and interaction can influence how an audience interprets the presented information. 
Hullman and Diakopoulos \cite{Hullman2011} formalized this observation by introducing the concept of visualization rhetoric to understand how design techniques in narrative visualizations prioritize particular interpretations and shape understanding.
The authors identify five classes of rhetorical strategies, each corresponding to a different editorial layer of a visualization. Together, these techniques capture how omissions, emphases, and framings can enter a visualization at multiple levels simultaneously. The detailed definitions of each rhetoric technique are provided in the Appendix; what follows is a high-level description. 

The first two strategies operate at the level of what information is made available and how trustworthy that information appears to be. 
\textit{Information Access Rhetoric} concerns decisions about which data to include, omit, or aggregate, shaping which patterns and narratives become visible to the viewer. 
\textit{Provenance Rhetoric}, by contrast, governs how a visualization signals its own trustworthiness through source citations, methodological transparency, or acknowledgment of uncertainty. 
These two strategies are closely related: as Prantl et al. noted \cite{Prantl2025}, signals of rigor and source transparency are not merely procedural norms but active rhetorical tools that can strengthen a visualization's emotional and persuasive appeal by making its claims feel believable \cite{Prantl2025}.
The remaining three strategies concern how information, once selected, is visualized and communicated. 
\textit{Mapping Rhetoric} describes the translation of data values into visual features such as position, size, scale, or color, a process that can create emphasis or distort perception beyond what the raw data supports. 
\textit{Linguistic-Based Rhetoric} operates through titles, labels, annotations, and captions, often employing metaphor, irony, and rhetorical questions to guide interpretation in ways that the visual encoding alone cannot define. 
Finally, \textit{Procedural Rhetoric} concerns the design of interactivity: default views, filters, and animations that guide users' navigation of the data and, consequently, the conclusions they are most likely to reach.

A key insight of Hullman and Diakopoulos is that these strategies rarely operate in isolation. 
They frequently co-occur across multiple editorial layers, reinforcing one another and making the overall framing effect stronger and less transparent than any single technique would be on its own. 
This interdependence is interesting while understanding misleading visualizations: a visualization may conform to conventional design standards at one layer while introducing significant distortion at another, making detection non-trivial even for an expert audience.


\section{Authorial Intents Behind Misleading Visualizations}
\label{sec:intents}

Multiple intentions may drive the author of a visualization, and effective use of data requires cognitive abilities and effort \cite{Kantor2025}. 
In pursuing these intents, the author can employ (intentionally or accidentally) various strategies that generate visual design violations or reasoning errors, producing a misleading visualization.
Assessing whether a visualization is misleading is a complex and often subjective task \cite{Lisnic2023,Yang2021}. 
Determining whether a visualization contains elements that could deceive a reader involves both identifying visual or reasoning errors and interpreting how different audiences might receive and understand the presented information. 
This subjectivity becomes even more evident when considering the author's intention to deceive through a misleading visualization. Determining whether an author deliberately intended to mislead or inadvertently produced a misleading visualization requires judgments that go beyond a simple analysis of common errors in the visualization.
Assessing the intentions behind a misleading visualization can help understand the ``\textit{why}'' behind a deceptive effect.

To address this challenge, we propose a novel taxonomy of author intents and contributing factors that may lead to misleading effects. This taxonomy defines nine intents, grouped into intentional and non-intentional categories. 
In general, the author's intent is often latent and rarely stated explicitly.
These intents are not mutually exclusive; multiple intents can simultaneously contribute to the deceptive effect.
By identifying the author's intent behind a misleading visualization, it becomes possible to reason more systematically about whether the misleading aspects are intentionally produced.

\para{Intentional Misleading Intents}

\begin{itemize}[leftmargin=1em]
\item \textit{Claim-Supporting Manipulation} The author shapes the visualization and its caption to highlight a desired pattern or hide counter-evidence. The goal is to persuade the reader to adopt a particular perspective, for example, by spreading disinformation about COVID-19 \cite{Lee2021,Soares2021,LisnicLex2024,Alexander2024,Dhawka2025,LoGupta2022,McNutt2020,Rho2025}.

\item \textit{Bias Exploitation} The visualization is crafted to take advantage of known human visual perceptual or cognitive biases in order to push the reader to a particular interpretation. For example, the author may use a truncated axis to amplify perceived trends \cite{LoGupta2022,Dimara2020,Correl2020,Yang2021,McNutt2020}.

\item \textit{Context Distortion} The author removes (or hides/omits) relevant context that is essential for accurate interpretation of the visualization (missing labels, explanations, legends). The goal is to persuade the reader towards a particular perspective \cite{McNutt2020,LoGupta2022,Mahbub2025,Hullman2011,Dhawka2025}.

\item \textit{Deliberate Reader Confusion} The author introduces unnecessary complexity, such as ambiguous encodings, to create confusion and make the reader more receptive to the author's narrative \cite{Lisnic2023,McNutt2021Villany,Borland2007,Lauer2020,Lauer2020HowPeople}.

\item \textit{Selective Reporting} The author reports only partial information, such as reporting partial data, focusing/highlighting on a data subset. The goal is to provide a skewed representation of the information, aiming at persuading the reader towards a particular perspective \cite{Hullman2011,Lisnic2023,Mahmud2017,Lauer2020HowPeople}.
\end{itemize}

\para{Unintentional Misleading Intents}

\begin{itemize}[leftmargin=1em]
\item \textit{Aesthetic-Driven Misrepresentation} The author prioritizes the visual appeal of the visualizations, giving them precedence over accuracy. These stylistic preferences, such as decorative choices, lead to misinterpretation, despite the author's intention not to mislead \cite{Bateman2010,McNutt2021Villany,Szafir2018,Borland2007}.

\item \textit{Lack of Visualization Literacy} The author is unfamiliar with best practices in visualization and inadvertently introduces misleading elements. For example, the author may choose inappropriate chart types or a design that inadvertently leads to misleading conclusions \cite{Lee2017Vlat,Varona2025,LoGupta2022}.

\item \textit{Space and Format Constraints} Constraints (such as space constraints, image size/format) limit the design or presentation of the visualization. To cope with these constraints, the author may incur information loss, which can lead to misrepresentation \cite{Hoffswell2020}.

\item \textit{Unintentional Context Omission} Accidental omissions of relevant context that are essential for accurate interpretation of the visualization (missing labels, explanations, legends). The misleading purpose is not deliberate \cite{McNutt2020,Ragan2016,Mahbub2025}. 

\end{itemize}

\section{Research Objectives}
\label{sec:objectives}

Misleading visualizations occur when design flaws distort the underlying data, preventing the reader from correctly interpreting it~\cite{McNutt2020,LoGupta2022,Lan2025,Lisnic2023,Lauer2020}.
Multiple intents may drive the author of a visualization. In pursuing these intents, the author can employ various rhetorical techniques (e.g., choices in data selection, visual encoding, and linguistic framing) that shape how the audience perceives the information. Depending on how these techniques are applied, the resulting visualization may incorporate visual design violations or reasoning errors, producing a misleading visualization.
Determining whether a visualization is misleading is both complex and often subjective~\cite{Lisnic2023,Yang2021}.
A key dimension of this complexity concerns intentionality.
A visualization may mislead because the author deliberately manipulated the design to support a particular narrative, or because of unintentional factors such as limited visualization literacy or aesthetic choices made without awareness of their distorting effect. 
Intentionality is inherently difficult to assess from visual artifacts alone, since the same design choice (e.g., a truncated axis) can arise either from deliberate manipulation or from accidental mistake. 
While recognizing rhetorical techniques involves identifying observable patterns in visual design, recognizing authorial intent requires deeper reasoning to infer the possible motivations behind a misleading visualization.

This paper aims to evaluate the capacity of multimodal LLMs to identify and interpret misleading visualizations along three directions. 
First, we examine whether these models can detect misleading elements under two distinct conditions: when asked to simply analyze and get insights from 1) a potentially misleading visualization, and 2) when explicitly informed that it has already been assessed as misleading. 
Second, we investigate whether multimodal LLMs can recognize the rhetorical techniques present in misleading visualizations, following the categories proposed by Hullman and Diakopoulos~\cite{Hullman2011}.
Third, we assess whether these models can infer authorial intentions or the contributing factors behind misleading effects, using the novel taxonomy introduced in \cref{sec:intents}. This third direction is the most demanding, as it requires the model not only to identify \textit{what} is wrong with a visualization but also to reason about \textit{why} it has been constructed that way.

\noindent To conduct this analysis, we formulate two primary research questions:

\begin{itemize}
    \item \RQuestion{1}: Are multimodal LLMs capable of identifying \textbfi{rhetorical techniques} in misleading data visualizations?
    \item \RQuestion{2}: Are multimodal LLMs capable of identifying \textbfi{authorial intents} behind misleading data visualizations?
\end{itemize}

\noindent Each question is investigated under three experimental conditions, defined by the degree of prior knowledge provided to the model:

\begin{itemize}[label=--] 
    \item \rcond{A} \ldots when \textbfi{not informed} that the visualization is misleading?
    \item \rcond{B} \ldots when \textbfi{explicitly informed} that the visualization is misleading?
    \item \rcond{C} \ldots when \textbfi{explicitly informed} that the visualization is misleading and provided with the \textbfi{specific misleading errors} it contains?
\end{itemize}

\noindent This generates six sub-questions that reflect two distinct types of analysis. Identifying rhetorical techniques is primarily an \textit{analytical} task: it requires the model to examine observable choices in the visualization's design and communication, and recognize how they shape interpretation. 
Identifying authorial intents is an \textit{inferential} task: it requires the model to reason beyond what is visible and attribute the observed design choices to possible motivations or circumstances, whether deliberate or accidental. 
Both dimensions are investigated across the same three levels of prior knowledge (\rcond{A}, \rcond{B}, \rcond{C}).

\noindent Finally, an additional question emerges naturally from the A conditions, in which the model is exposed to both misleading and non-misleading visualizations without any prior warning. Since no indication of potential deception is provided, any judgment about whether a visualization is misleading must arise entirely from the model's own critical reasoning:

\begin{itemize}
    \item \rquestion{0}: Can multimodal LLMs \textbfi{identify} misleading visualizations, without being explicitly instructed to look for predefined deceptive elements? When they do, what elements do they attend to, and how do they characterize the misleading nature of the visualization?
\end{itemize}


\subsection{RQ0 -- Identify Misleading Visualizations}

This question asks whether LLMs can identify misleading visualizations.
It emerges naturally from the A experimental conditions.
Since the goal is to provide no warning or instruction to look for particular deceptive elements, any misleading judgment must arise entirely from the model's own \textit{unprompted} critical reasoning.
This setting is deliberately minimal, and its minimalism is what makes it meaningful. It simulates the most realistic end-user scenario: a non-expert who asks an LLM to help interpret a visualization they encounter, without any prior suspicion of possible manipulation. In this sense, \rquestion{0} reflects the conditions under which misleading visualizations are most likely to cause damage, that is, when the viewer, human or model, has no reason to be on guard.

This distinguishes our approach from prior work such as Alexander et al.~\cite{Alexander2024}, who explicitly instructed the model to identify misleading aspects and provided definitions of errors to discover as part of the prompting.
In their setup, the detection task is \textit{targeted}: the model is told what to look for, and its role is to confirm or deny its presence. In our \rcond{A} conditions, the task is \textit{open-ended}: the model must decide on its own whether the visualization is misleading and, if so, discuss why.
In addition, this distinguishes our approach from Lo and Qu \cite{Lo2025}, who explored more complex prompting strategies (e.g., few-shot learning, chain-of-thought \cite{Wei2022}) to assess the model's ability to detect misleading visualizations.

This framing of the research question also motivates the second part of \rquestion{0}. We are interested not only in whether the model detects the misleading visualization, but also in what elements it attends to and how it characterizes the misleading effect. The answer to this question reveals what the model \textit{spontaneously} considers problematic in a visualization, independent of any predefined taxonomy of errors.

\subsection{RQ1 -- Identify Rhetorical Techniques}

This research question investigates whether LLMs can recognize the rhetorical techniques that shape and influence interpretation in misleading visualizations.
As detailed in \cref{sec:rhetoric_def}, rhetorical techniques are the communicative strategies by which a visualization influences the viewer's reading, regardless of whether it contains explicit errors.
These strategies include choices about which data to include or omit, how visual encodings are selected to emphasize certain patterns, and how linguistic elements, such as titles and captions, are used to frame the narrative.

The three sub-questions test this capability under increasing levels of prior knowledge, but the progression is not only additive; each condition also changes the nature of the task the model is asked to perform.
In \rquestion{1A}, the model receives no prior information and must first decide whether the visualization is misleading. Rhetorical identification is, in some way, \textit{spontaneous}: it can occur only if the model first reaches a positive, misleading judgment, and both the detection and the rhetorical reasoning must emerge entirely from the model's own \textit{unprompted} critical analysis.
In \rquestion{1B}, the model is told that the visualization is misleading, eliminating the detection step entirely. The task becomes a \textit{rhetorical identification}: the model knows something is wrong and must locate and characterize the communicative strategies responsible for the misleading effect.
In \rquestion{1C}, the model is additionally provided with the specific errors that made the visualization misleading. This shifts the task from identification to \textit{anchored explanation}: the errors are already known, and the model must reason about how they connect to, and are expressed through, broader rhetorical strategies.


\subsection{RQ2 -- Identify Authorial Intents}

This research question investigates whether LLMs can identify the authorial intents or contributing factors behind misleading visualizations.
As detailed in \cref{sec:intents}, authorial intents refer to the underlying motivations or circumstances (deliberate and/or accidental) that may lead a visualization to mislead.
These range from intentional strategies, such as exploiting cognitive biases or selectively reporting data to support a narrative, to unintentional factors, such as limited visualization literacy or aesthetic choices that inadvertently compromise the visualization accuracy.

Compared to \rquestion{1}, this question requires deeper reasoning from the model. Identifying rhetorical techniques requires the model to recognize observable design patterns, an analytical task based on what is directly visible. 
Identifying authorial intents requires the model to go further, inferring the motivations that may have driven those design choices: this is an inferential task that requires reasoning about what lies behind the actual visualization. 
Furthermore, this inference must account for the ambiguity of intentionality: the same design choice can arise from deliberate manipulation or by mistake, and the model must reason about which is more plausible given the available evidence.

The three sub-questions test this capability under the same progression of prior knowledge as RQ1, with each condition again changing the nature of the task.
The range from \textit{spontaneous} attribution (\rquestion{2A}), through \textit{identification}~(\rquestion{2B}), to \textit{anchored explanation}~(\rquestion{2C}).
The key difference is that the model must now reason about motivations and circumstances rather than observable communicative strategies.

\section{Study Design}
\label{sec:study}

To address the research questions, we designed the following study. It is organized around three main components: a dataset of COVID-19 social media posts containing misleading and non-misleading visualizations (\cref{sec:dataset}), six experiments testing the LLMs under varying levels of prior knowledge (\cref{sec:experiments,sec:models}), and a user study with visualization experts who annotated the same visualizations (\cref{sec:userstudy}).
A schematic representation of the study is shown in \cref{fig:teaser}.


\subsection{Models}
\label{sec:models}

To ensure broad coverage of the current landscape of multimodal LLMs, we selected 16 state-of-the-art models. 
Among them, 15 are open-weight models spanning a wide range of model sizes, architectural families, and reasoning capabilities, while the sixteenth, GPT-5.4 (OpenAI), is a proprietary frontier model included as a competitive reference point. 
Our goal was to evaluate the research questions across models designed to run on a range of hardware, from lightweight, consumer-grade devices to systems requiring specialized GPU infrastructure, with a primary focus on open-weight models that enable transparent replication and comparative analysis. 
The selection includes both dense architectures and Mixture-of-Experts (MoE) models, several of which feature reasoning capabilities.
As shown in \cref{tab:models}, the 15 open-weight models are organized into three groups of five based on total parameter count: \textit{small} ($\leq$32B), \textit{medium} (70--124B), and \textit{large} ($\geq$235B). 
This grouping reflects the hardware requirements for deployment: small models target consumer-grade single-GPU setups, medium models typically require multi-GPU configurations, and large models demand specialized high-memory infrastructure. GPT-5.4 is reported separately as its architecture and parameter count are undisclosed. For readability, we assign each model a short nickname (e.g., \deepseek, \gemma) used consistently throughout the paper. \cref{tab:models-detail} provides additional details.
All open-weight models were executed using their official vendor-provided weights and in the default (highest-precision) format. When multiple model sizes were available, we selected the largest-parameter variant. Due to hardware limitations, \maverick\ and \stepthree\ were executed using the vendor-provided \texttt{FP8} quantized variants. GPT-5.4 was accessed through the OpenAI API. Furthermore, we used the default temperature provided by each model.

\begin{table}[t]
\centering
\caption{
The 16 LLMs evaluated in this study. 
The \textit{ID} column provides the nickname used throughout the paper. 
Models are organized into three groups based on total parameter count.
All models are open-weight except for \gpt. Additional information about these models are in \cref{tab:models-detail}.
}
\vspace{-2mm}
\label{tab:models}
\small
\setlength{\tabcolsep}{5pt}
\begin{tabular}{@{}lcllrr@{}}
\toprule

\textbf{ID} & & \textbf{Model} & \textbf{Provider} & \textbf{Params (B)} & \\

\midrule
\multicolumn{6}{@{}l}{\textit{Small models (P\,$\leq$ 33\,B)}} \\[2pt]
\nemotron & \cite{nvidia2025nvidianemotronnanov2} & Nemotron-Nano-V2-VL & Nvidia & 12 & \hflink{nvidia/NVIDIA-Nemotron-Nano-12B-v2-VL-BF16}\\
\mistral & & Mistral-Small-3.2 & Mistral AI & 24 & \hflink{mistralai/Mistral-Small-3.2-24B-Instruct-2506}\\
\deepseek & \cite{wu2024deepseekvl2} & DeepSeek-VL2 & DeepSeek & 27 & \hflink{deepseek-ai/deepseek-vl2}\\
\gemma & \cite{gemmateam2025gemma3technicalreport} & Gemma3 & Google & 27 & \hflink{google/gemma-3-27b-it}\\
\gta & \cite{yang2025gta1guitesttimescaling} & GTA1 & Salesforce & 32 & \hflink{Salesforce/GTA1-32B}\\

\midrule
\multicolumn{6}{@{}l}{\textit{Medium models (70--124\,B total)}} \\[2pt]
\qianfan & \cite{qianfan-vl-2025} & Qianfan-VL & Baidu & 70 & \hflink{baidu/Qianfan-VL-70B}\\
\molmo & \cite{deitke2024molmopixmoopenweights} & Molmo & Ai2 & 72 & \hflink{allenai/Molmo-72B-0924}\\
\glm & \cite{vteam2025glm45} & GLM-4.5V & Z.ai & 108 & \hflink{zai-org/GLM-4.5V}\\
\llava & \cite{li2024llavanext} & LLaVA-NeXT & LLaVA & 110 & \hflink{llava-hf/llava-next-110b-hf}\\
\pixtral & & Pixtral-Large & Mistral AI & 124 & \hflink{mistralai/Pixtral-Large-Instruct-2411}\\

\midrule
\multicolumn{6}{@{}l}{\textit{Large models (Total $\geq$ 235\,B)}} \\[2pt]
\qwen & \cite{qwen3technicalreport} & Qwen3-VL & Alibaba & 235 & \hflink{Qwen/Qwen3-VL-235B-A22B-Instruct}\\
\intern & \cite{wang2025internvl} & InternVL3.5 & OpenGVLab & 241 & \hflink{OpenGVLab/InternVL3_5-241B-A28B}\\
\stepthree & \cite{step3system} & Step3 & StepFun AI & 321 & \hflink{stepfun-ai/step3-fp8}\\
\maverick & & Llama-4-Maverick & Meta & 400 & \hflink{meta-llama/Llama-4-Maverick-17B-128E-Instruct-FP8} \\
\kimi & \cite{kimiteam2026kimik25} & Kimi-K2.5 & Moonshot AI & 1,000 & \hflink{moonshotai/Kimi-K2.5} \\
\midrule
\gpt & & GPT-5.4-2026-03-05 & OpenAI & -- & \\

\bottomrule
\end{tabular}
\vspace{-3mm}
\end{table}
\subsection{Dataset}
\label{sec:dataset}

This study analyzes visualizations shared on social media during the COVID-19 pandemic, a period marked by intense public debate, uncertainty, and the widespread dissemination of data-driven narratives. 
COVID-19, being a highly polarized and emotionally charged topic, has pushed users to share data (even misleading) visualizations that support their personal beliefs, challenge public health measures, or reinforce their political positions \cite{Alexander2024}. As a result, many misleading visualizations were produced that leverage both rhetorical strategies and authorial intent.
To support our experiments, we extracted
a subset from the dataset collected by Lisnic et al. \cite{Lisnic2023}. The original dataset includes 9,958 English-language posts shared on \textit{X} (formerly Twitter) during the COVID-19 pandemic, each containing at least one data visualization. Lisnic et al. annotated these tweets to indicate whether the visualization was misleading. For misleading cases, the authors also specified the type of visualization design violation (e.g., \textit{truncated axis}, \textit{unclear encoding}) and/or reasoning error (e.g., \textit{cherry-picking}, \textit{causal inference}) that contributed to the misinterpretation of data.
Among them, 2,373 tweets contain at least one misleading element, while the remaining 7,585 were categorized as not misleading. 

However, due to X’s content policies, Lisnic et al. shared only tweet IDs rather than the full content or images. To construct a balanced dataset, we first attempted to retrieve all 2,373 misleading tweets. 
We successfully collected 1,168 misleading tweets; the rest were unavailable due to removal, privacy settings, or technical issues. We then collected a random sample of 1,168 non-misleading tweets. In total, our dataset contains \textbf{2,336 tweets} (50\% misleading, 50\% not misleading). 

\subsection{Experiments with LLMs}
\label{sec:experiments}

We designed six experiments organized along two orthogonal dimensions: \emph{topic} ($x$) and \emph{experimental condition} ($y$). Each experiment \experiment{\textit{xy}} addresses \rquestion{\textit{xy}} using prompt \prompt{\textit{xy}}, where $x \in \{1, 2\}$ and $y \in \{\text{A}, \text{B}, \text{C}\}$. Experiments \experiment{1y} target rhetorical techniques; \experiment{2y} target authorial intents. All 16 models were tested in each experiment. An overview is shown in \cref{fig:teaser}.

\para{Prompts}
Each prompt follows a common template with three tasks (\cref{fig:prompt}; exact text in the supplemental material). In \textit{task~1}, the model analyzes the visualization and its caption: it describes what the visualization represents, interprets its message, and draws its own conclusion. \textit{Task~2} concerns misleading aspects and differs across conditions (see below). \textit{Task~3} asks the model to identify which rhetorical techniques (\prompt{1x}) or authorial intents (\prompt{2x}) contributed to the misleading effect, explain how each contributes, and rate each on a scale from $0$ (\textit{none}) to $6$ (\textit{very strong}), with $-1$ if unsure. All models produce structured JSON output, available as supplemental material.
The three conditions control what the model knows before tasks~2 and~3. In condition \rcond{A}, the model receives no information about misleadingness and must independently determine whether the visualization is misleading; rhetoric or intent identification proceeds only if it concludes that it is. This makes the task \emph{spontaneous}. In condition \rcond{B}, the model is explicitly told that the visualization is misleading, removing the detection step and turning the task into focused \emph{identification}. In condition \rcond{C}, the model additionally receives the ground-truth error labels, shifting the task to \emph{anchored explanation}: the errors are known, and the model may reason about how they connect to rhetorical strategies or authorial intents.

Condition \rcond{A} involves all 2,336 visualizations (50\% misleading, 50\% non-misleading) and also addresses \rquestion{0}. Conditions \rcond{B} and \rcond{C} involve only the 1,168 misleading visualizations.

\begin{figure}[t]
\vspace{-2mm}
\centering
\fbox{\includegraphics[width=0.95\linewidth]{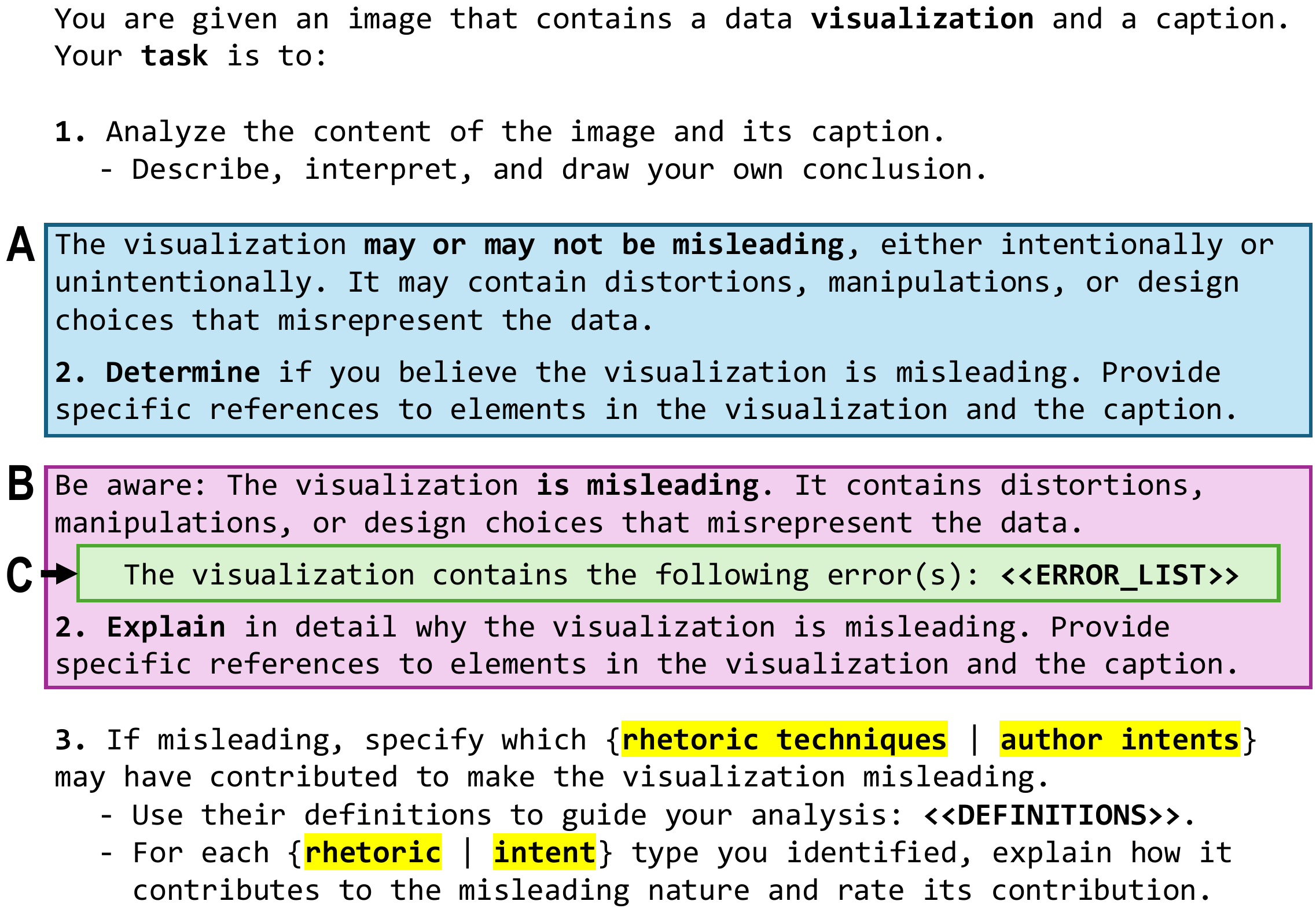}}
\vspace{-2mm}
\caption{
Schematic representation of the prompt template.
Black text is shared across all conditions.
Condition \cbA\ does not inform the model about misleadingness.
Condition \cbB\ states that the visualization is misleading.
Condition \cbC\ extends \rcond{B} by providing the ground-truth error list.
Task~3 asks the model to identify rhetorics (\experiment{1x}) or intents (\experiment{2x}).
}
\label{fig:prompt}
\end{figure}

\para{Experiment Design}
The six experiments form a $2 \times 3$ design. The two topics differ in the type of reasoning they demand. 
Experiments \experiment{1A}, \experiment{1B}, and \experiment{1C} ask models to identify the five rhetorical techniques (\cref{sec:rhetoric_def}) that shape interpretation in misleading visualizations. 
This is an \emph{analytical} task: it requires recognizing observable patterns concerning, for example, visual design, data selection, and linguistic framing. 
Experiments \experiment{2A}, \experiment{2B}, and \experiment{2C} ask models to identify the nine authorial intents (\cref{sec:intents}) behind the misleading effect. 
This is an \emph{inferential} task: it requires reasoning beyond what is visible to attribute the observed design choices to possible motivations, whether deliberate or accidental. 
The three conditions (\rcond{A}, \rcond{B}, \rcond{C}) are identical across the two topics, and the prompts differ only in whether they request rhetoric or intent attribution in task~3.
\subsection{User Study with Visualization Experts}
\label{sec:userstudy}

To provide a human reference for the rhetoric and intent assessment tasks, we conducted a user study with visualization experts, who assessed the same misleading visualizations under a condition equivalent to \rcond{C}: each visualization was presented together with its ground-truth error label. We chose visualization experts because assessing misleading visualizations requires advanced competencies in data visualization, media interpretation, and scientific reasoning \cite{Alexander2024}.

\para{Participants} 
A total of 11 visualization experts participated: 3 Associate Professors, 2 Assistant Professors, 3 Postdoctoral Researchers, and 3 PhD Students. Self-assessed expertise in misleading visualization analysis averaged 4.45 (SD~$=1.21$) on a 7-point scale; expertise in visualization rhetoric averaged 3.64 (SD~$=1.63$).

\para{Design}
Presenting all 1,168 misleading visualizations to each participant would be unrealistic, so we designed a focused evaluation strategy. We decomposed the dataset into \textit{\textless error, visualization\textgreater} pairs, associating each error type with the visualizations in which it appears. Among the 14 error types, two were extremely rare (\textit{Incorrect reading of the chart}: 4 instances; \textit{Uneven binning}: 3) and were removed. Each participant analyzed one visualization per remaining error type, resulting in 12 visualizations per expert. To avoid ordering effects, we randomized both the sequence of errors and the assigned visualization using a Latin square design. The study was conducted using the Revisit platform \cite{revisit1, revisit2}.

\para{Assessments}
For each \textit{\textless error, visualization\textgreater} pair, participants viewed the original tweet, its visualization, and the labeled error from the ground-truth annotations. 
To maintain context and mirror the LLM experimental setup, in which definitions were included directly in the prompt, we complemented each assessment with the full definitions of the error types, rhetoric techniques (\cref{sec:rhetoric_def}), and authorial intents (\cref{sec:intents}).
Participants rated the contribution of each of the five rhetorical categories and each of the nine intent categories using the same scale as the LLMs.

\section{Results}
\label{sec:results}

This section presents the results organized by research question. We first introduce the analytical framework used to evaluate model behavior in \rquestion{1} and \rquestion{2} (\cref{sec:framework}), then report findings for each research question. 
Examples of LLM outputs are provided in the supplemental material; additionally, an interactive web-based explorer for navigating model responses, embeddings, and attribution profiles, along with complete results, is available at {\small\url{https://github.com/XAIber-lab/truevislies}}.

\subsection{Analytical Framework}
\label{sec:framework}

In all six experiments, models rated the contribution of each rhetoric category or authorial intent to the misleading nature of a visualization on a scale from $0$ (\textit{none}) to $6$ (\textit{very strong}), with $-1$ indicating uncertainty. These ratings allow us to characterize how models distribute attribution across the error types present in the dataset, independently of whether the ratings were produced spontaneously (\rcond{A}), under informed conditions (\rcond{B}), or anchored to ground-truth errors (\rcond{C}). From these ratings, we derive three analytical quantities.

\para{Contribution Probability Matrix} For each model $m$ at a given experimental condition, the \textit{mean score matrix} $\mu_m(r,\,e)$ records the mean contribution score assigned to category $r$ (rhetoric or intent) across all tweets containing error $e$. Normalizing each column by its sum yields the \textit{contribution probability matrix}, whose entries represent the conditional probability $P(r \mid e)$ of a category given an error type, independently of overall scoring level.

\para{Error Sensitivity Score (ESS)} This measures how strongly a model's attribution behavior varies across error types. For each category $r$, ESS is the standard deviation of its row in the mean score matrix: $\sigma_m[\,\mu(r,\,e)\,]$, then averaged over all categories to produce a single scalar. A high ESS indicates that the model assigns systematically different profiles to different errors; a low ESS indicates approximately uniform attribution regardless of error type. ESS is a sensitivity measure, not a quality measure: the ideal model should approach the human ESS level. A model can have low ESS for the right reason (recognizing genuine co-occurrence) or for the wrong reason (assigning flat, uninformative scores). ESS shares its conceptual foundation with variance-based global sensitivity analysis~\cite{Saltelli2002,Sobol2021}; similar approaches have been used to evaluate prompt sensitivity~\cite{Zhuo2024,Sclar2024}.

\para{Model Behavioral Similarity (MBS)} For each model $m$, a behavioral vector is constructed by concatenating the flattened mean score matrix with the flattened standard deviation matrix (the latter scaled by $\frac{1}{3}$). Each cell is row-normalized by subtracting the row mean. Pairwise cosine similarity between behavioral vectors measures whether two models activate the same categories for the same errors with comparable consistency.

\subsection{Expected Results}
\label{sec:expected_results}
The 12 error types in the dataset~\cite{Lisnic2023} belong to two families. \textit{Visualization design violations} operate through the graphical encoding of data values (e.g., truncated axes, value as area), while \textit{reasoning errors} operate through the selection and framing of evidence (e.g., cherry-picking, causal inference). Each family should activate a distinct attribution profile.
For rhetoric, \textit{Mapping} should dominate for design violations, while \textit{Information Access} and \textit{Linguistic-Based} rhetoric should rise for reasoning errors. \textit{Provenance} should remain at a stable, moderate level across both families. \textit{Procedural} rhetoric should be negligible given the static nature of tweet visualizations. A theoretically consistent model should therefore produce a matrix in which the design/reasoning boundary corresponds to a visible transition from Mapping-dominant to Information Access and Linguistic-dominant columns.
For intent, \textit{intentional} categories (Claim-Supporting Manipulation, Selective Reporting, Context Distortion, Deliberate Reader Confusion, Bias Exploitation) should show higher probabilities for reasoning errors, while among \textit{unintentional} categories, Aesthetic-Driven Misrepresentation and Lack of Visualization Literacy should be elevated for design violations. Unintentional Context Omission and Space/Format Constraints should remain low and evenly distributed.

\subsection{RQ0 – Identify Misleading Visualizations}
\label{sec:rq0_results}

Addressing \rquestion{0} focuses on the binary classification results from \experiment{1A} and \experiment{2A}, in which LLMs determined whether each visualization was misleading without being told to look for any particular deception. Given the balanced dataset, \textit{accuracy} serves as an interpretable baseline. We additionally report the \textit{Matthews Correlation Coefficient} (MCC), which provides a more reliable summary by accounting for all four confusion-matrix outcomes~\cite{Chicco2020, Chicco2023}. Unlike F1-score, MCC is not biased toward the positive class: it ranges in $[-1,1]$, where $+1$ denotes perfect agreement, $0$ indicates chance-level performance, and $-1$ total disagreement. We base our analysis primarily on MCC; accuracy and F1 are reported in the appendix (\cref{fig:app-accuracy}).

\begin{figure}[t]
\vspace{-2mm}
\centering
\includegraphics[width=\linewidth]{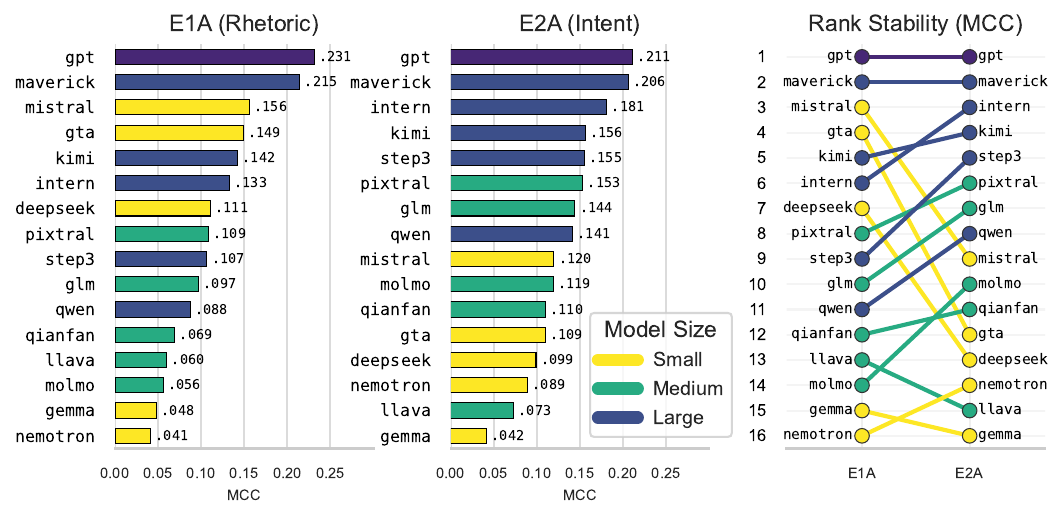}
\vspace{-2mm}
\caption{
MCC scores for all 16 models in \experiment{1A} (Rhetoric) and \experiment{2A} (Intent), with rank stability between the two conditions shown on the right. Overall performance is low in both experiments (mean MCC $0.113$ and $0.132$, respectively), with \gpt\ and \maverick\ leading consistently. Color encodes model size (small, medium, large); larger models tend to score higher, a trend that is statistically significant in \experiment{2A}.
}
\label{fig:mcc}
\vspace{-2mm}
\end{figure}

Performance across both experiments is overall low (\cref{fig:mcc}). In \experiment{1A}, MCC ranges from $0.041$ (\nemotron) to $0.231$ (\gpt). 
In \experiment{2A}, MCC ranges from $0.042$ (\gemma) to $0.211$ (\gpt). Model rank ordering is broadly consistent between the two experiments but not identical: \gpt\ and \maverick\ lead in both, all medium and large models improve ranking except \llava, while all small models drop except \nemotron. A Spearman rank correlation between total parameters and MCC for \experiment{2A} is significant ($\rho=0.83$, $p < 0.001$); in \experiment{1A} the trend is present but does not reach significance. These results suggest that scale matters more when the task demands inferential reasoning about intent (\experiment{2A}) than when it requires identifying observable rhetorical patterns (\experiment{1A}). However, scale alone is not sufficient: within the large group, \maverick\ (MoE, 17B active from 400B total) and \intern\ (241B, 28B active) substantially outperform \kimi\ (1T total, 32B active), demonstrating that active parameter count and architectural design mediate classification quality.

The Wilcoxon signed-rank test reveals a significant difference in MCC between \experiment{1A} and \experiment{2A} ($W=27.0$, $p=0.034$), confirmed by a paired t-test ($t=-2.14$, $p=0.049$). Models achieve slightly higher MCC under the intent-framing prompt (mean $0.132$) than the rhetoric-framing prompt (mean $0.113$), suggesting that reasoning about authorial intent generates slightly more selective classifications despite similar raw accuracy. The effect is small, and both experiments remain in the weak-discrimination range; nonetheless, it indicates that prompt framing influences discriminative quality.

A common pattern across both experiments is a tendency to classify visualizations as misleading regardless of the ground truth. For most models, recall substantially exceeds precision. \gemma\ flags $99.7\%$ of all visualizations as misleading (recall~$=0.997$, precision~$=0.502$). Even \gpt\ achieves a recall of $0.863$ while precision is only $0.564$. This bias means that high accuracy is achieved despite, rather than because of, meaningful discrimination: most models approximate a degenerate classifier that labels almost everything as misleading. Only \maverick, \gta, and \llava\ show more balanced precision-recall profiles (recall~$\approx 0.50$).

\begin{figure*}[ht]
    \vspace{-2mm}
    \centering

    \begin{subfigure}[c]{0.48\linewidth}
        \centering
        \includegraphics[width=\linewidth]{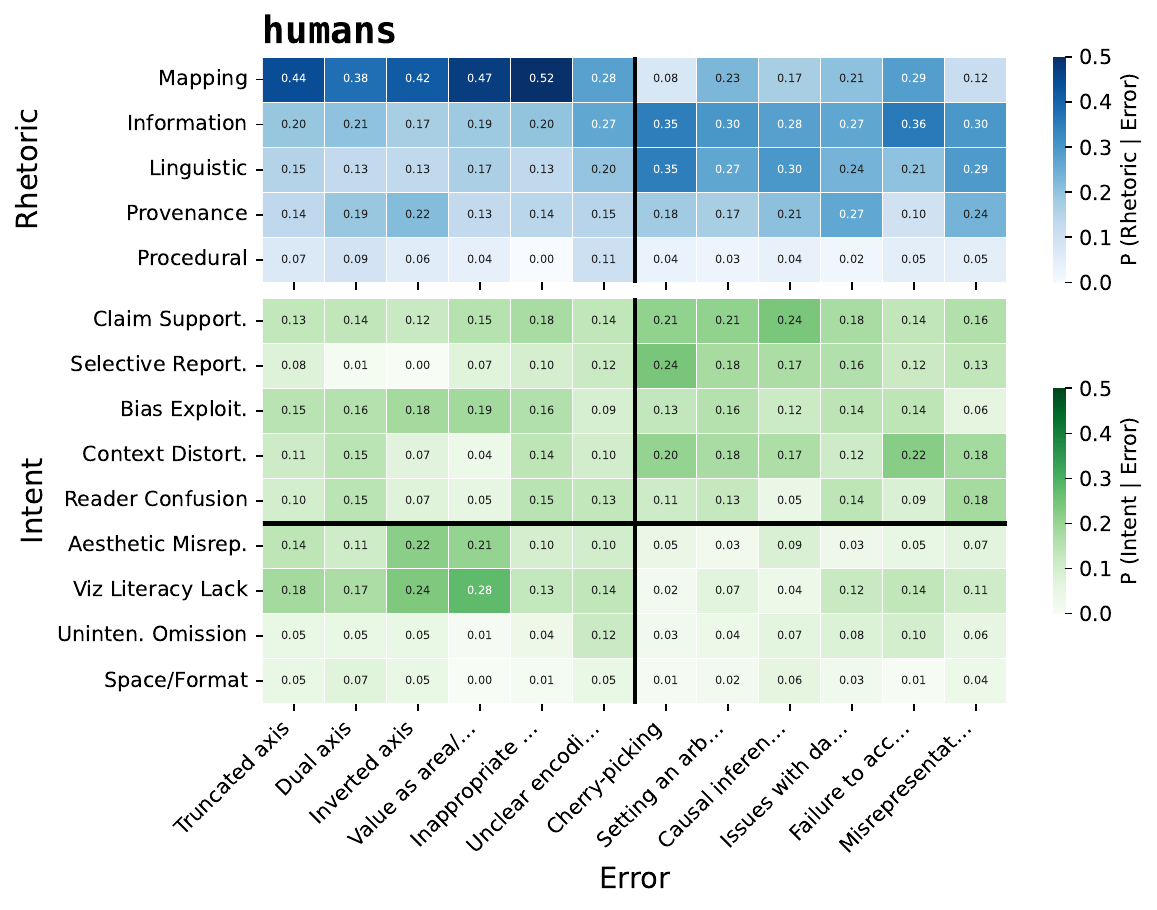}
        \caption{Humans}
        \label{fig:p_rhetoric_intent_error_human}
    \end{subfigure}
    \hfill
    \begin{subfigure}[c]{0.5\linewidth}
        \centering

        \begin{subfigure}[t]{0.48\linewidth}
            \centering
            \includegraphics[width=\linewidth]{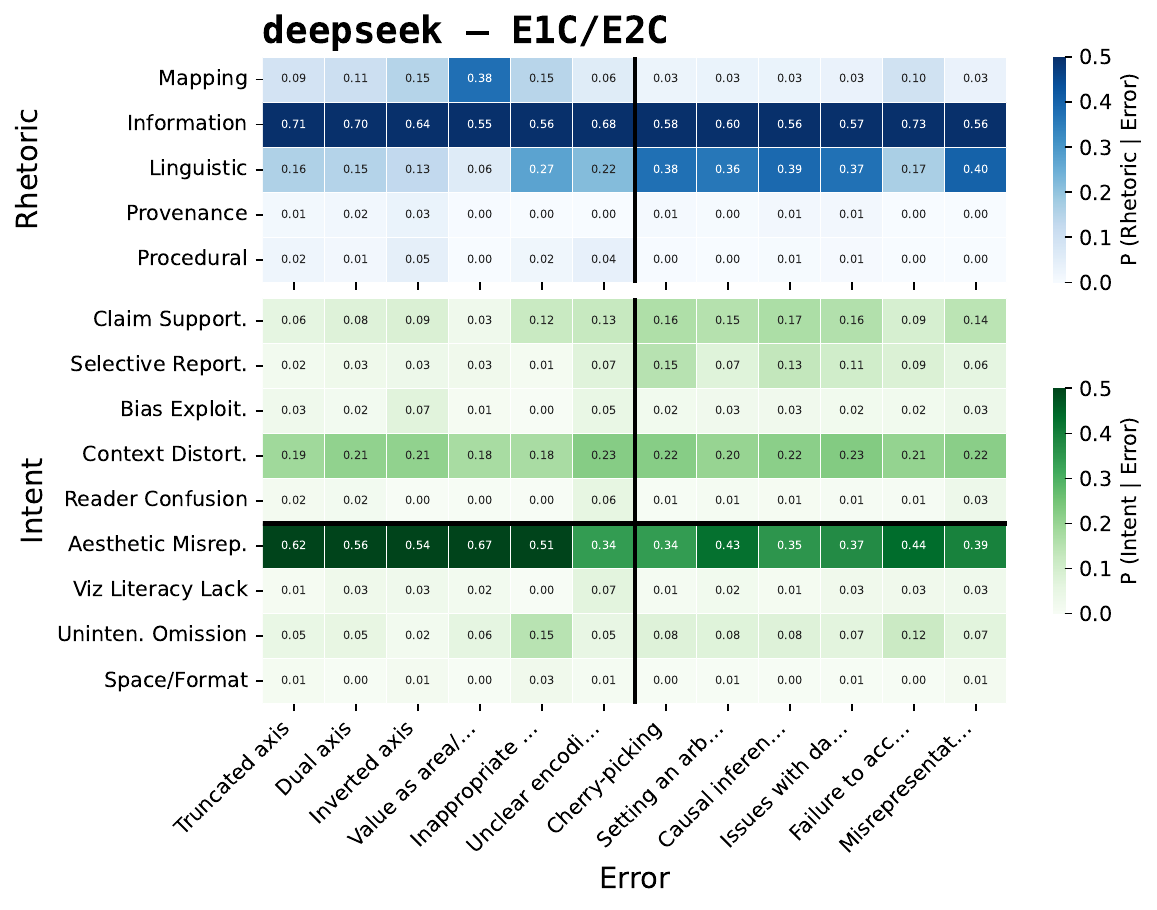}
            \caption{\deepseek}
            \label{fig:p_rhetoric_intent_error_deepseek}
        \end{subfigure}
        \hfill
        \begin{subfigure}[t]{0.48\linewidth}
            \centering
            \includegraphics[width=\linewidth]{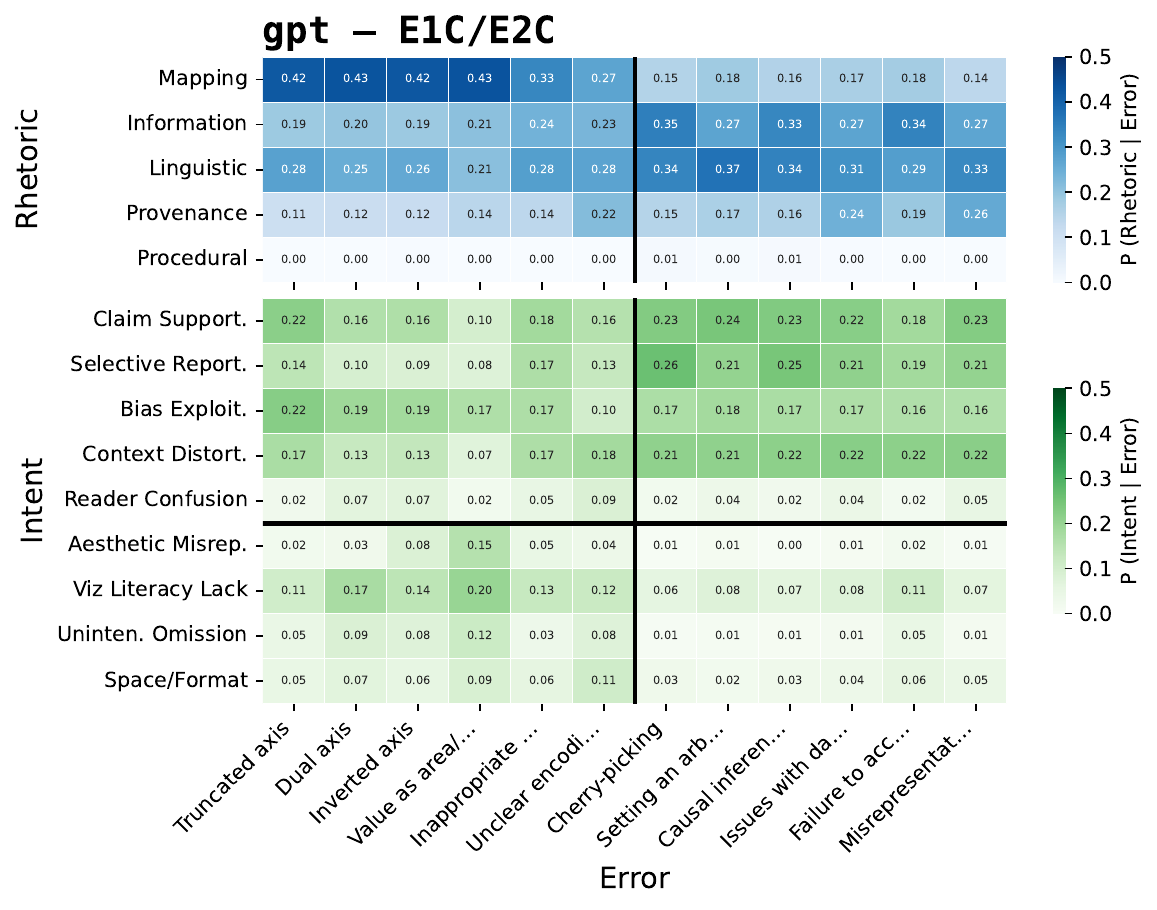}
            \caption{\gpt}
            \label{fig:p_rhetoric_intent_error_gpt}
        \end{subfigure}

        \begin{subfigure}[t]{0.48\linewidth}
            \centering
            \includegraphics[width=\linewidth]{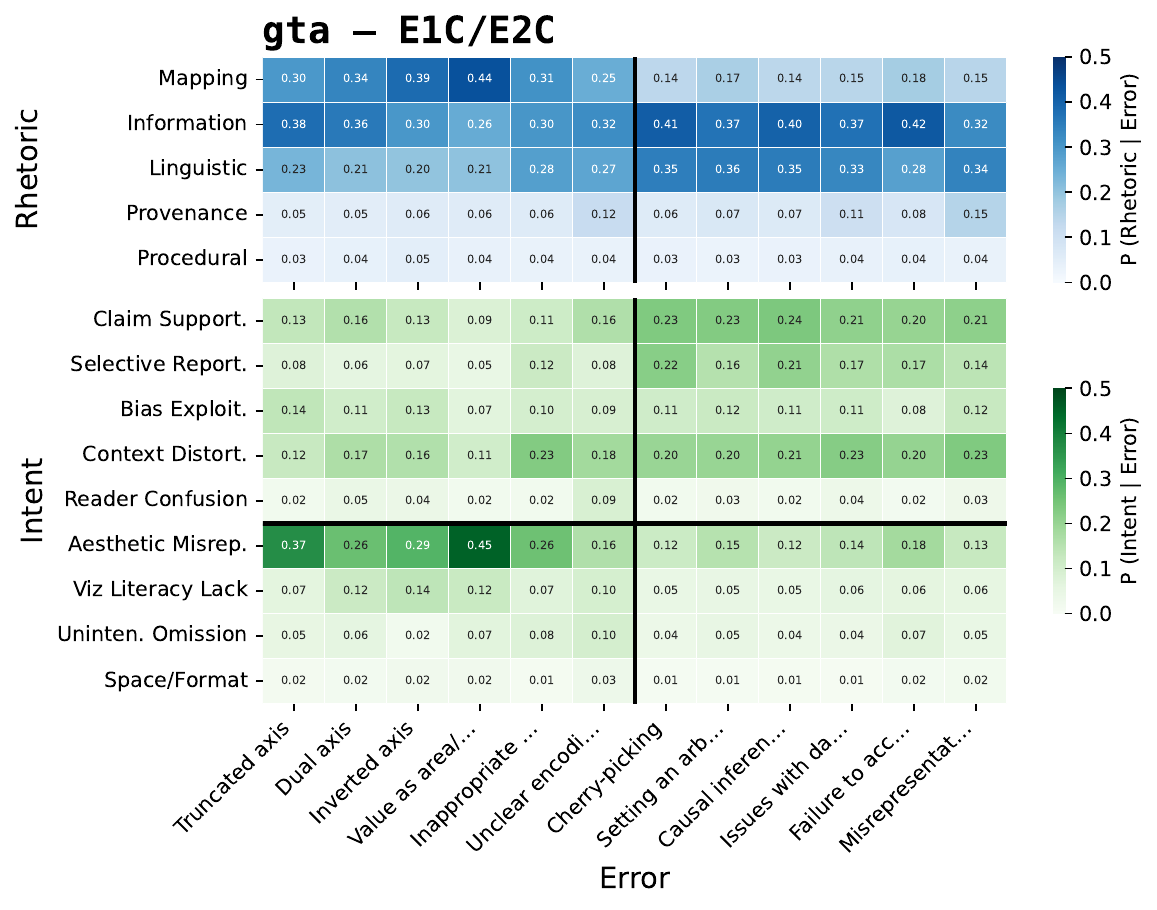}
            \caption{\gta}
            \label{fig:p_rhetoric_intent_error_gta}
        \end{subfigure}
        \hfill
        \begin{subfigure}[t]{0.48\linewidth}
            \centering
            \includegraphics[width=\linewidth]{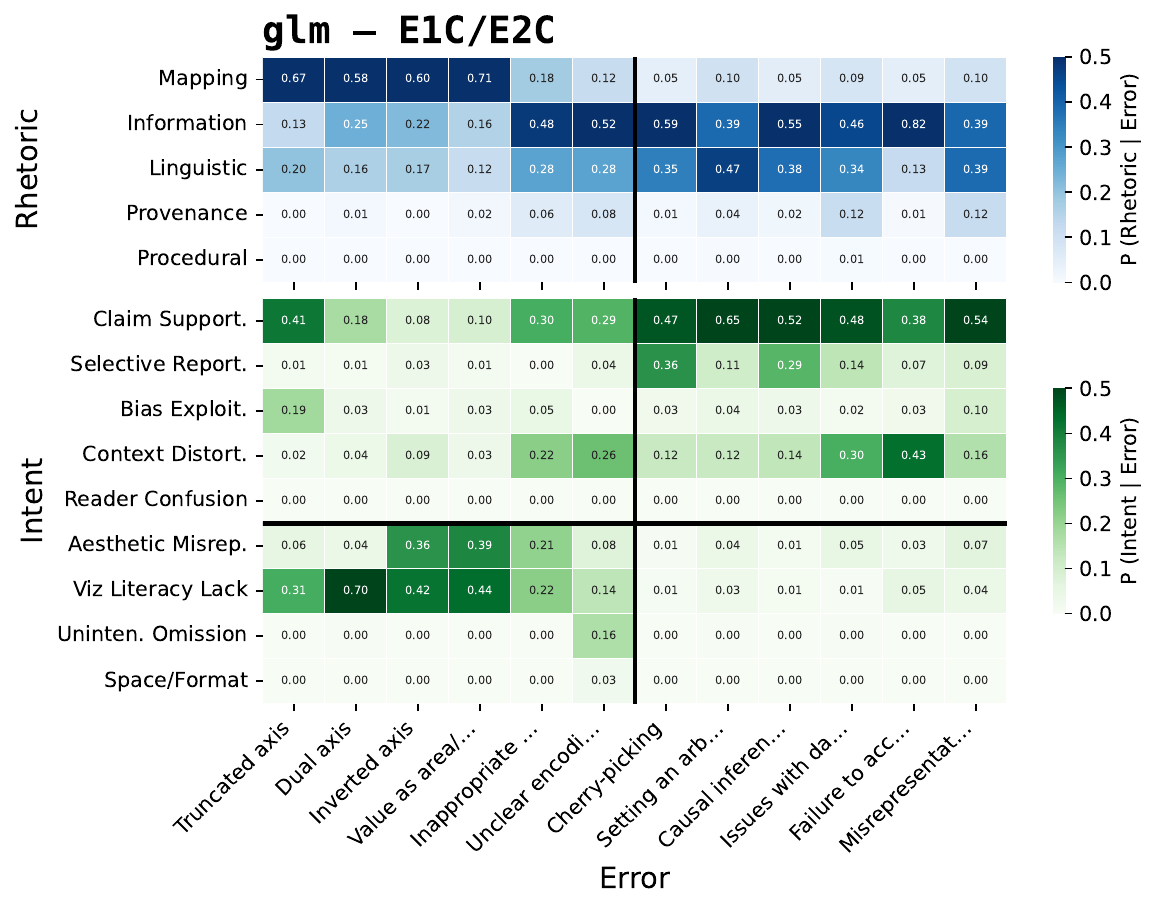}
            \caption{\glm}
            \label{fig:p_rhetoric_intent_error_glm}
        \end{subfigure}
    \end{subfigure}
    \vspace{-4mm}
    \caption{
Contribution probability matrices $P(\text{rhetoric} \mid \text{error})$ (blue) and $P(\text{intent} \mid \text{error})$ (green) collected from the visualization experts (a) and selected models (b--e) under experimental condition \rcond{C}.
Columns correspond to error types; the vertical black line separates \textit{visualization design violations} (left) from \textit{reasoning errors} (right).
For rhetoric by humans~(a), the cross-family transition is visible: \textit{Mapping} dominates the design violation columns, while \textit{Information Access} and \textit{Linguistic} rise for reasoning errors.
For intent by humans~(a), the horizontal black line separates intentional (top) from unintentional (bottom) intents; unintentional intents have higher probabilities for design violations, and intentional intents for reasoning errors.}
    \label{fig:p_rhetoric_intent_error}
\end{figure*}

\para{Semantic Analysis}
To qualitatively characterize LLMs' explanations, we embed the models' textual responses using \texttt{Qwen3-Embedding-8B} \cite{qwen3embedding} --- a state-of-the-art embedding model --- and project them into two dimensions with UMAP. The resulting projections (shown in \cref{fig:teaser} and detailed in \cref{fig:umap_whymis}) form a single, densely packed cloud with no visible class boundary. The silhouette scores are $0.055$ (\experiment{1A}) and $0.042$ (\experiment{2A}), both near zero, confirming that model explanations for truly misleading and non-misleading visualizations are globally entangled in the semantic space. Models produce semantically similar explanations regardless of whether the visualization is genuinely misleading. Truly misleading tweets elicit approximately $1.6\times$ more varied explanations than non-misleading ones, consistent with the diversity of error types in the dataset. A preliminary BERTopic clustering further reveals that the primary organizing principle of the embedding space is the subject of the visualization (e.g., country-level COVID statistics) rather than the type of misleading error, indicating that model explanations are anchored in visual content more than in an abstract concept of misleadingness.

\para{Summary} All models operate in the weak-discrimination range (MCC $0.04$--$0.23$). \gpt\ and \maverick\ lead consistently; most models exhibit a strong positive-labeling bias. Scale correlates significantly with MCC for the inferential intent framing but not the analytical rhetoric framing. Model explanations are semantically entangled regardless of ground truth, organized by visualization topic rather than error type.

\subsection{RQ1 -- Identify Rhetorical Techniques}
\label{sec:rq1_results}
To characterize how LLMs assess rhetorical techniques across error types, we apply the analytical framework presented in \cref{sec:framework}.

\para{Human Baseline}
Experts annotated misleading visualizations under a condition equivalent to \rcond{C} and rated each rhetoric category using the same scale as the models. The resulting contribution probability matrix (\cref{fig:p_rhetoric_intent_error_human}) follows the expected pattern closely (\cref{sec:expected_results}). For visualization design violations, Mapping rhetoric dominates, reaching $0.52$ for Inappropriate encoding and remaining above $0.38$ for all other design errors except one. For reasoning errors, Mapping drops while Information Access and Linguistic-Based rhetoric rise. Provenance maintains stable values across both families, with a slight elevation for Misrepresentation of scientific studies ($0.24$), where source credibility is directly implicated. Procedural is negligible. Experts produce high ESS ($0.910$), meaning they sharply differentiate rhetoric attributions across error types. This is the expected behavior: a visualization expert should assign a very different rhetoric profile to a truncated axis than to a cherry-picked visualization.

\para{Model Attribution Profiles}
The dominant failure mode in conditions \rcond{A} and \rcond{B} is the over-attribution of Information Access rhetoric, which most models apply uniformly across all 12 error types regardless of error family, functioning as a catch-all explanation for misleadingness. This is misaligned with the expert baseline: Information Access should rise specifically for reasoning errors, not for design violations. The failure is most extreme in \deepseek\ ($0.57$--$0.74$ in \rcond{A}, persisting at $0.55$--$0.73$ through \rcond{C}; \cref{fig:p_rhetoric_intent_error_deepseek}), and is present to varying degrees in \gemma, \intern, and \gta\ (\cref{fig:p_rhetoric_intent_error_gta}), all of which show no cross-family differentiation across any condition. \glm\ (\cref{fig:p_rhetoric_intent_error_glm}) shows a distinct failure mode: per-error instability in which Information Access and Provenance alternate dominance unpredictably, with no coherent family-level structure.
Among the models that develop a cross-family structure in \rcond{C}, \gpt\ (\cref{fig:p_rhetoric_intent_error_gpt}) and \maverick\ stand apart. In conditions \rcond{A} and \rcond{B}, both maintain balanced rhetoric profiles without collapsing onto a single dominant category. In condition \rcond{C}, both develop the clearest and most coherent cross-family differentiation in the set.

\para{Error Sensitivity}
ESS results are shown in \cref{fig:ess_rhetoric}. The most notable pattern is a massive jump from conditions \rcond{A}/\rcond{B} to \rcond{C}: ESS values approximately span $0.25$--$0.54$ in \experiment{1A} and \experiment{1B}, but reach $0.38$--$1.01$ in \experiment{1C}. Providing the ground-truth error roughly doubles sensitivity for most models, confirming that the error acts as a strong discriminative hook. Within conditions \rcond{A} and \rcond{B}, \kimi, \gpt, and \maverick\ lead. In \experiment{1C}, the ranking reshuffles: \mistral\ moves to 1st, suggesting it is particularly effective at leveraging error information to sharpen attributions. The bottom tier is stable across all conditions: \llava, \deepseek, and \gta\ persistently occupy the last positions, suggesting a limited ability to modulate attribution by error type regardless of prompting.

A three-way ANOVA on ESS is highly significant ($F(2,45)=36.99$, $p<0.001$). The \rcond{A} to \rcond{B} drop is small but significant ($\Delta=0.04$, $p<0.001$ Wilcoxon); the \rcond{A} to \rcond{C} and \rcond{B} to \rcond{C} jumps are large (both $\Delta \approx 0.35$, $p<0.001$). Condition \rcond{C} is the primary driver of sensitivity; the misleading-only restriction (\rcond{A} vs. \rcond{B}) has a minor effect.

A Spearman rank correlation between total parameters and ESS is negative, reaching significance in \experiment{1B} ($\rho=-0.54$, $p=0.032$): larger models tend to produce more uniform, less discriminating attribution profiles. With the human reference at $0.910$, small and medium models like \mistral\ and \glm\ exceed it in \experiment{1C}, while several large models (\intern, \qwen) remain well below. Providing the error partially closes this gap: the negative correlation weakens from \experiment{1B} ($\rho=-0.54$) to \experiment{1C} ($\rho=-0.35$, not significant), suggesting the error information acts as an equalizer. Even so, bottom-tier models \llava\ and \deepseek\ remain far below the human reference.

\begin{figure}[t]
    \vspace{-2mm}
    \centering
    \begin{subfigure}[t]{0.48\linewidth}
        \centering
        \includegraphics[width=\linewidth]{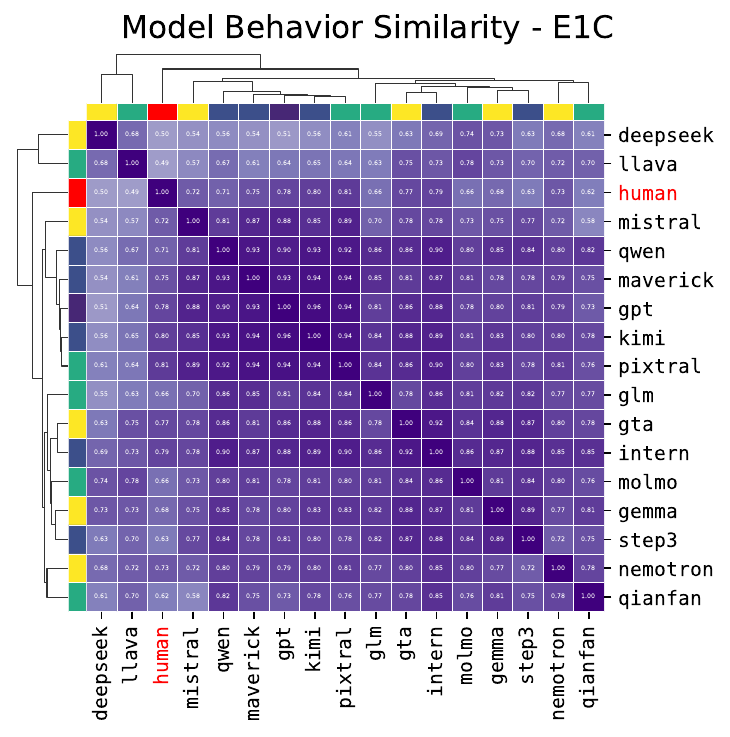}
        \caption{Rhetoric (\experiment{1C})}
        \label{fig:similarity_rhetoric_E1C}
    \end{subfigure}
    \hfill
    \begin{subfigure}[t]{0.48\linewidth}
        \centering
        \includegraphics[width=\linewidth]{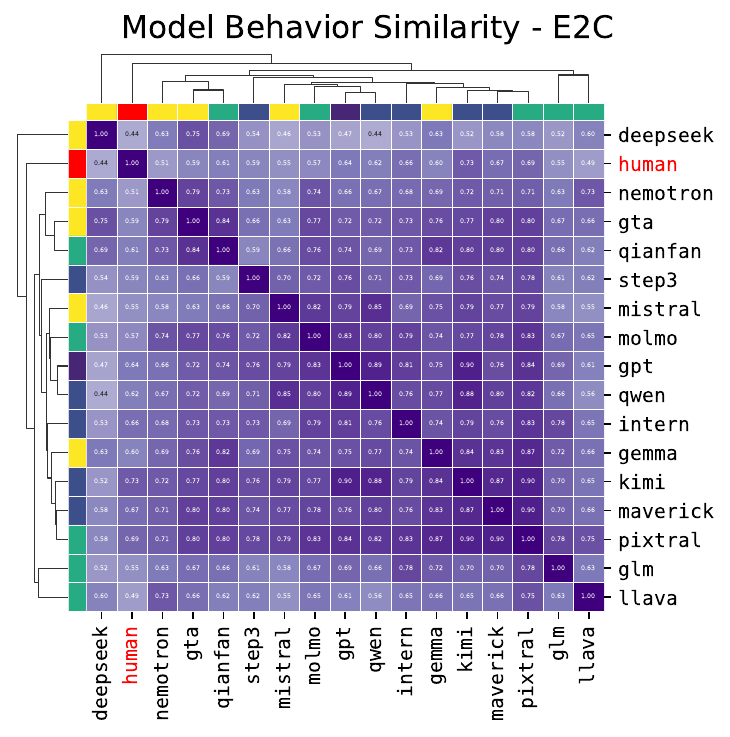}
        \caption{Intent (\experiment{2C})}
        \label{fig:similarity_intent_E2C}
    \end{subfigure}
    \vspace{-4mm}
    \caption{%
        Pairwise behavioral similarity matrices under condition~\rcond{C} for rhetoric~(a) and intent~(b).
        Rows and columns are ordered by hierarchical clustering.
        By condition~\rcond{C}, providing the ground-truth error drives models toward a common behavior, compressing the similarity range relative to conditions~\rcond{A} and~\rcond{B} (see \cref{fig:similarity_rhetoric_app,fig:similarity_intent_app}).
        \llava\ and \deepseek\ remain the most isolated models; \gpt\ and \kimi\ form the most stable high-similarity pair.
        Human similarity with the model population remains moderate (at most $0.81$ for rhetoric, $0.73$ for intent), confirming that models converge toward a shared yet human-divergent pattern.
    }
    \label{fig:similarity_rhetoric_intent}
\end{figure}

\para{Behavioral Similarity}
\cref{fig:similarity_rhetoric_intent} (left) shows the pairwise behavioral similarity matrix for rhetoric in \experiment{1C}; full matrices for \experiment{1A} and \experiment{1B} are in \cref{fig:similarity_rhetoric_app}. The most prominent trend across conditions is a progressive compression of pairwise similarity, indicating that providing the ground-truth error drives models toward a common discriminative behavior. \llava\ and \deepseek\ are the most isolated models across all conditions, consistent with the ESS findings. \gpt\ and \kimi\ form the most stable high-similarity pair, indicating structural alignment in how they relate error types to rhetoric categories. Human similarity with the model population remains moderate (at most $0.81$), confirming that models converge to a common but non-human behavior.

\begin{figure}[t]
    \vspace{-2mm}
    \centering
    \begin{subfigure}{0.3\linewidth}
        \centering
        \includegraphics[width=\linewidth]{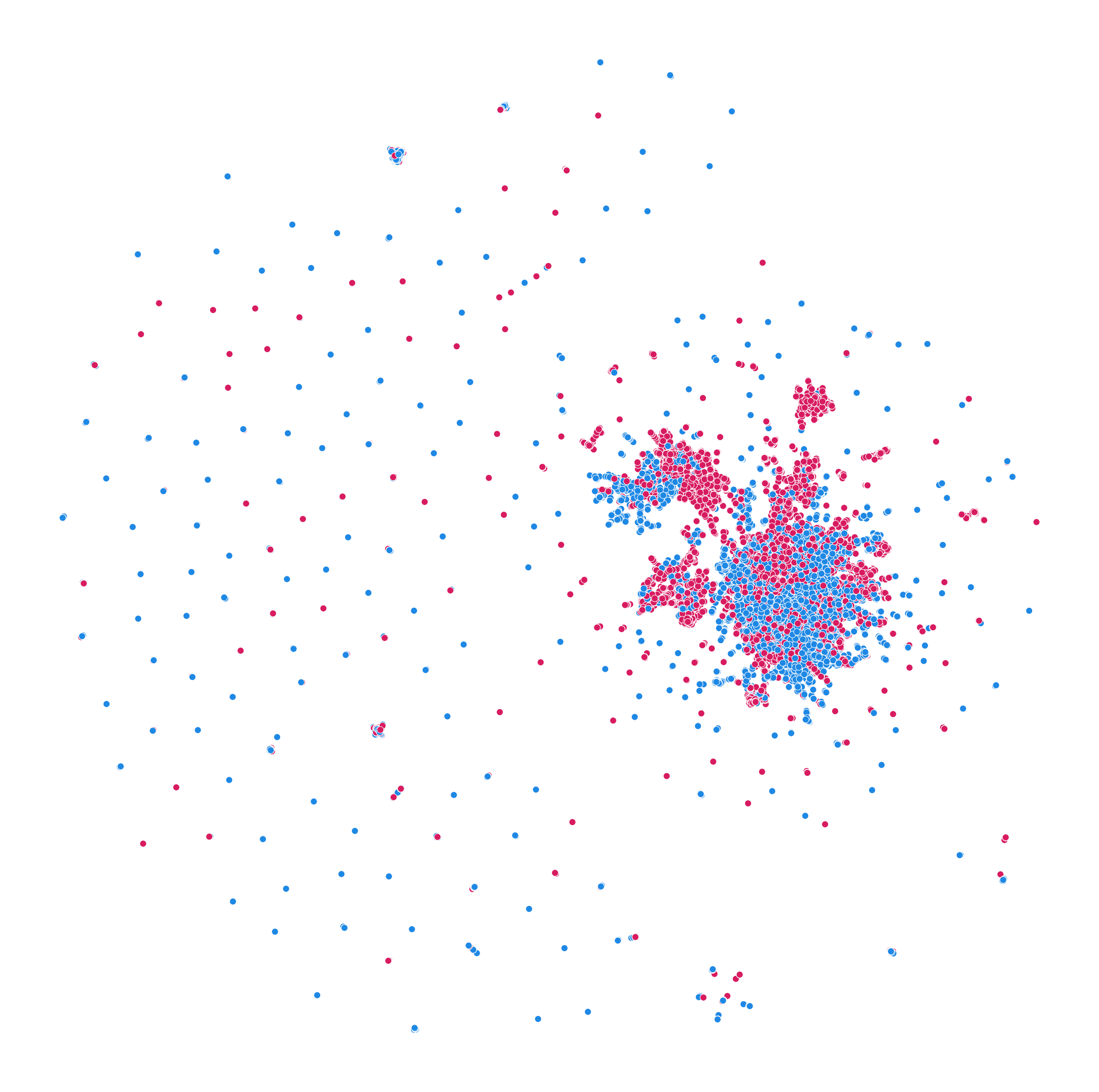}
        \caption{\experiment{1A}}
    \end{subfigure}
    \hfill
    \begin{subfigure}{0.3\linewidth}
        \centering
        \includegraphics[width=\linewidth]{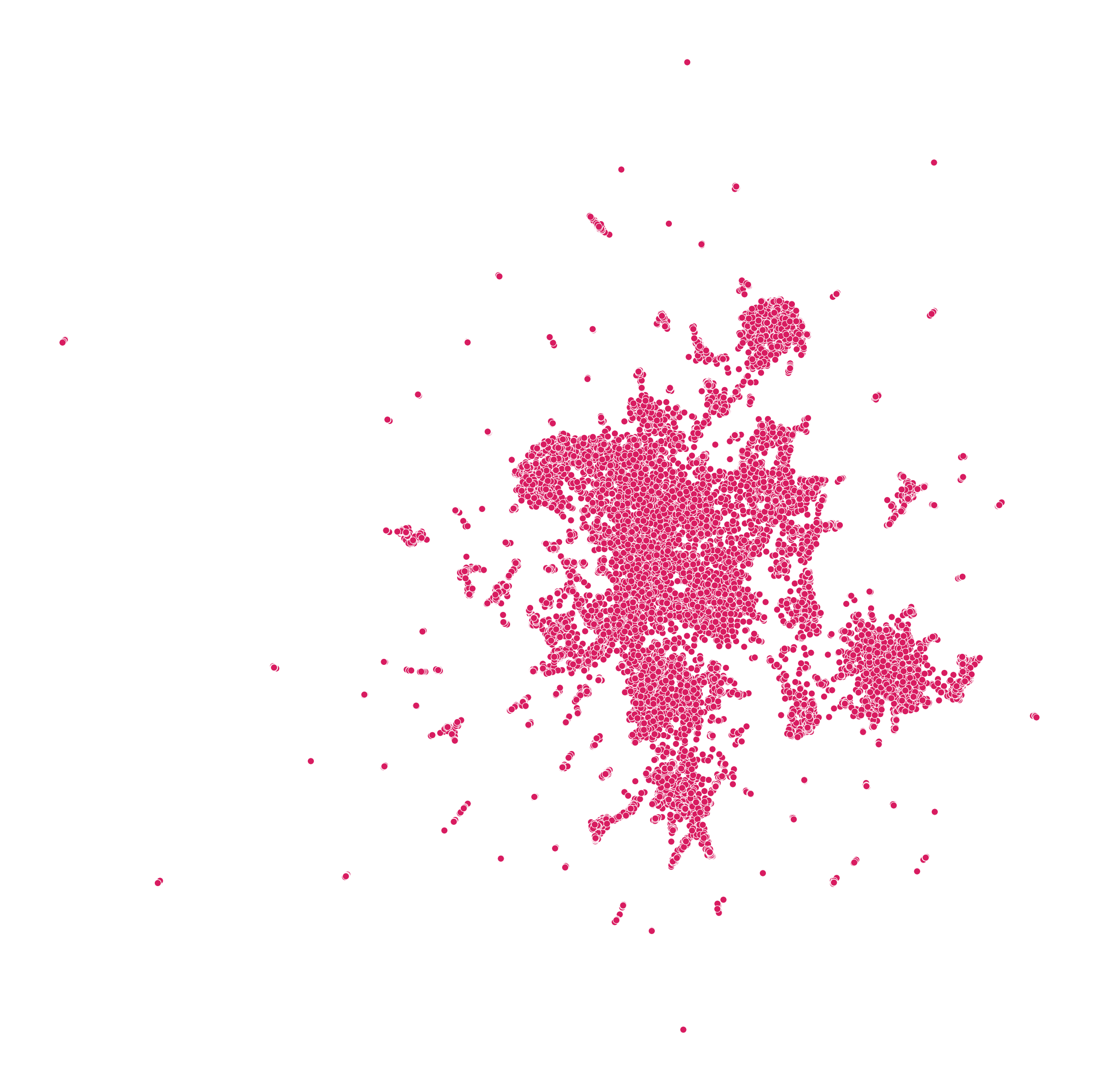}
        \caption{\experiment{1B}}
    \end{subfigure}
    \begin{subfigure}{0.3\linewidth}
        \centering
        \includegraphics[width=\linewidth]{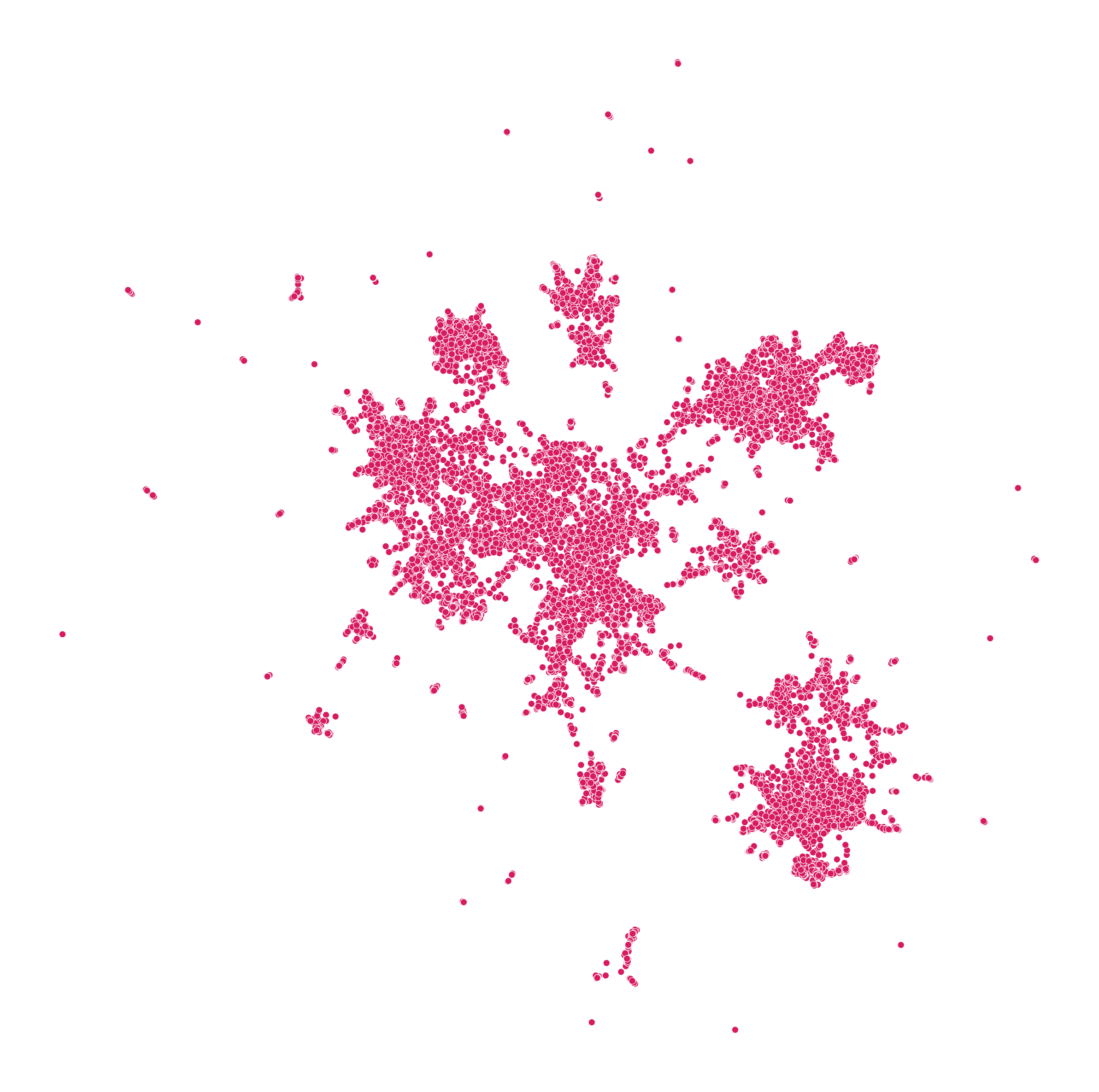}
        \caption{\experiment{1C}}
    \end{subfigure}
    \vspace{-2mm}
    \caption{UMAP projections of rhetoric contribution explanations across experiments \experiment{1A}, \experiment{1B}, and \experiment{1C}. Each point represents one tweet-model pair; color encodes ground truth (blue: non-misleading, red: misleading). The progressive dispersion of the embedding space reflects the increasing prior knowledge provided to the model across conditions.}
    \label{fig:umap_rhetoric}
\end{figure}

\para{Semantic Analysis}
UMAP projections of rhetoric contribution explanations (\cref{fig:umap_rhetoric}) show that the semantic space becomes progressively more dispersed across conditions \experiment{1A}--\experiment{1B}--\experiment{1C}. When models receive no contextual information beyond the image, rhetorical explanations collapse into a narrow semantic region; informing the model that the tweet is misleading and further anchoring it on the specific error type progressively unlocks a wider and more differentiated space of rhetorical reasoning.

\para{Summary} The dominant failure mode is uniform over-attribution of Information Access rhetoric regardless of error family. Only \gpt\ and \maverick\ develop the theoretically expected cross-family differentiation in condition~\rcond{C}. Providing ground-truth errors roughly doubles ESS, but models converge toward a shared non-human behavioral pattern. Larger models tend to produce less-discriminating attribution profiles, though error information partially closes this gap.

\subsection{RQ2 -- Identify Authorial Intents}
\label{sec:rq2_results}

To characterize how LLMs assess authorial intents across error types, we apply the same analytical framework as for rhetoric (\cref{sec:framework}), substituting the nine intent categories for the five rhetoric categories.

\para{Human Baseline}
The intent probabilities collected from the visualization experts (\cref{fig:p_rhetoric_intent_error_human}) follow the expected cross-family pattern (\cref{sec:expected_results}). Among intentional categories, Claim-Supporting Manipulation is consistently high for reasoning errors, peaking at $0.24$ for Causal inference, while Selective Reporting shows the sharpest single-error spike at $0.24$ for Cherry-picking and drops to near zero for design violations. Context Distortion is stable in the reasoning half; Bias Exploitation and Deliberate Reader Confusion are weakly discriminating across both families. Among unintentional categories, Lack of Visualization Literacy carries the most informative signal, elevated for design violations ($0.13$--$0.28$) and receding for reasoning errors ($0.02$--$0.14$). Experts produced ESS $=0.880$, lower than for rhetoric ($0.910$), consistent with the larger number of categories (9 vs. 5), leading to a more balanced distribution.

\para{Model Attribution Profiles}
Condition \rcond{C} reveals three separable signals, none fully aligned with expert judgment across all models. Aesthetic-Driven Misrepresentation is the most correctly oriented unintentional category: \gta\ produces the clearest cross-family differentiation (\cref{fig:p_rhetoric_intent_error_gta}), closely matching the direction and magnitude of the experts' pattern. \gpt\ (\cref{fig:p_rhetoric_intent_error_gpt}) shows a weaker but directionally correct signal for Lack of Visualization Literacy, which remains near zero for reasoning errors, while its Aesthetic-Driven Misrepresentation profile is nearly flat. \deepseek\ (\cref{fig:p_rhetoric_intent_error_deepseek}) is the primary outlier, assigning the highest value regardless of error type. Overall, no model simultaneously replicates the expert cross-family structure across all intent categories: \gta\ comes closest for Aesthetic-Driven Misrepresentation, \gpt\ for Bias Exploitation, but both fail on Selective Reporting.

\para{Error Sensitivity}
ESS results for intent (\cref{fig:ess_intent}) mirror the rhetoric pattern: a large jump from conditions \rcond{A}/\rcond{B} (range $0.21$--$0.63$) to \rcond{C} (up to $0.93$). Unlike rhetoric, the top of the ranking is remarkably stable: \kimi\ and \gpt\ hold the first two positions in every experiment, suggesting a robust capacity for intent-level discrimination. In \experiment{2C}, \maverick\ rises to 3rd, consistent with its strong cross-family profiles. The bottom tier is equally stable: \deepseek\ is last in all three conditions.

A notable difference from rhetoric is the absence of the rank reshuffling observed in \experiment{1C}. For rhetoric, \mistral\ surged from the middle to 1st in condition \rcond{C}; for intent, condition \rcond{C} reinforces existing differences rather than creating new ones. This may reflect the greater inferential difficulty of intent attribution: leveraging the error taxonomy to sharpen intent requires reasoning about latent authorial motivations, so models that lack this capacity in \rcond{A}/\rcond{B} do not acquire it from the taxonomy alone.

The three-way ANOVA on intent ESS is highly significant ($F(2,45)=25.73$, $p<0.001$). The \rcond{A} to \rcond{B} comparison is not significant, unlike rhetoric where it produced a small but significant drop. This suggests that intent attribution, being inferential rather than analytical, does not benefit from the contrast between misleading and non-misleading cases. The \rcond{A} to \rcond{C} and \rcond{B} to \rcond{C} jumps remain the dominant effect ($\Delta \approx 0.29$, $p<0.001$), though the gain is slightly smaller than for rhetoric, consistent with the additional reasoning step from error to latent motivation.

The Spearman correlation between total parameters and ESS is negative and stronger than for rhetoric, reaching significance in \experiment{2C} ($\rho=-0.66$, $p=0.005$). Unlike rhetoric, where the negative correlation weakened from \experiment{1B} to \experiment{1C}, here it strengthens: the error hooks do not act as an equalizer for intent, and larger models remain less sensitive even when given explicit error information. With experts at ESS~$=0.88$, only \kimi\ exceeds the human reference in \experiment{2C}.

\para{Behavioral Similarity}
\cref{fig:similarity_rhetoric_intent} (right) shows the pairwise similarity matrix for intent in \experiment{2C}; the full set across conditions is in \cref{fig:similarity_intent_app}. The overall convergence tendency mirrors rhetoric but is less compact, consistent with the inferential nature of intent attribution. In \experiment{2A} the landscape is fragmented, with \llava, \gta, and \deepseek\ isolated. \deepseek\ becomes more dissimilar under condition \rcond{B} rather than converging, indicating that the instruction to assume misleadingness destabilizes its intent profile. By \rcond{C}, models converge, but humans remain moderately distant ($0.44$--$0.73$), confirming that models converge toward a shared yet human-divergent pattern.

\begin{figure}[t]
    \centering
    \begin{subfigure}{0.3\linewidth}
        \centering
        \includegraphics[width=\linewidth]{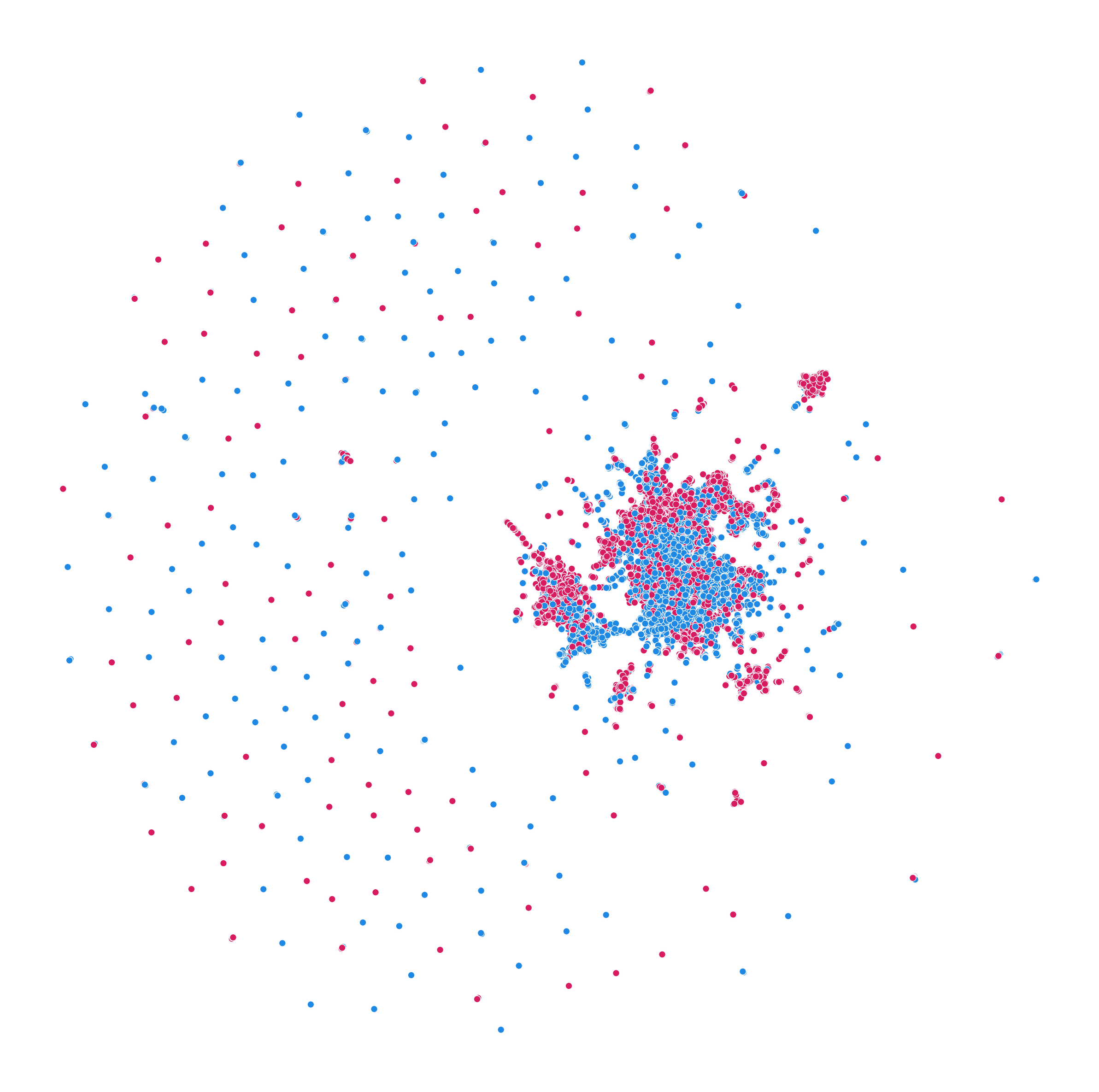}
        \caption{\experiment{2A}}
    \end{subfigure}
    \hfill
    \begin{subfigure}{0.3\linewidth}
        \centering
        \includegraphics[width=\linewidth]{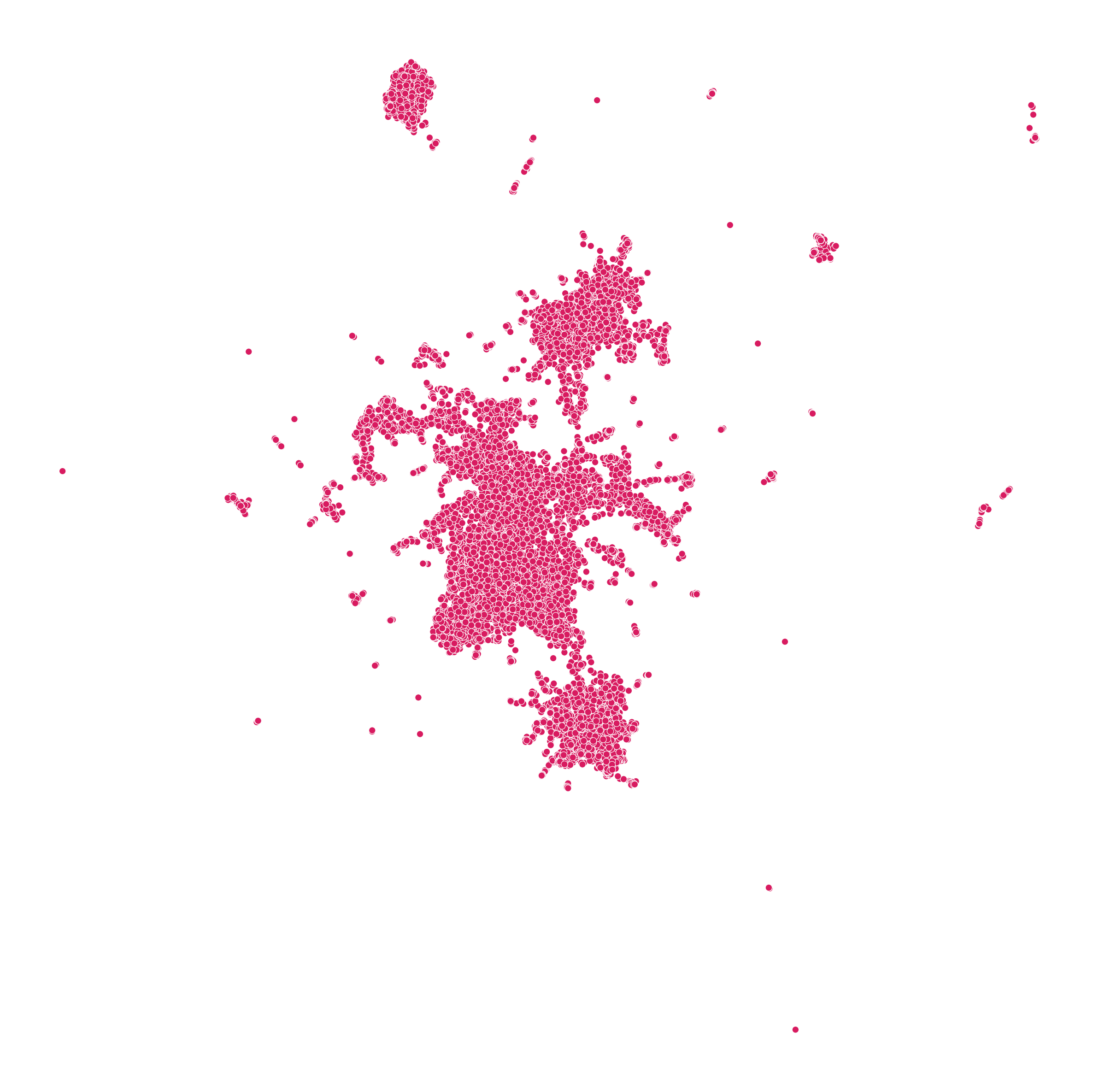}
        \caption{\experiment{2B}}
    \end{subfigure}
    \begin{subfigure}{0.3\linewidth}
        \centering
        \includegraphics[width=\linewidth]{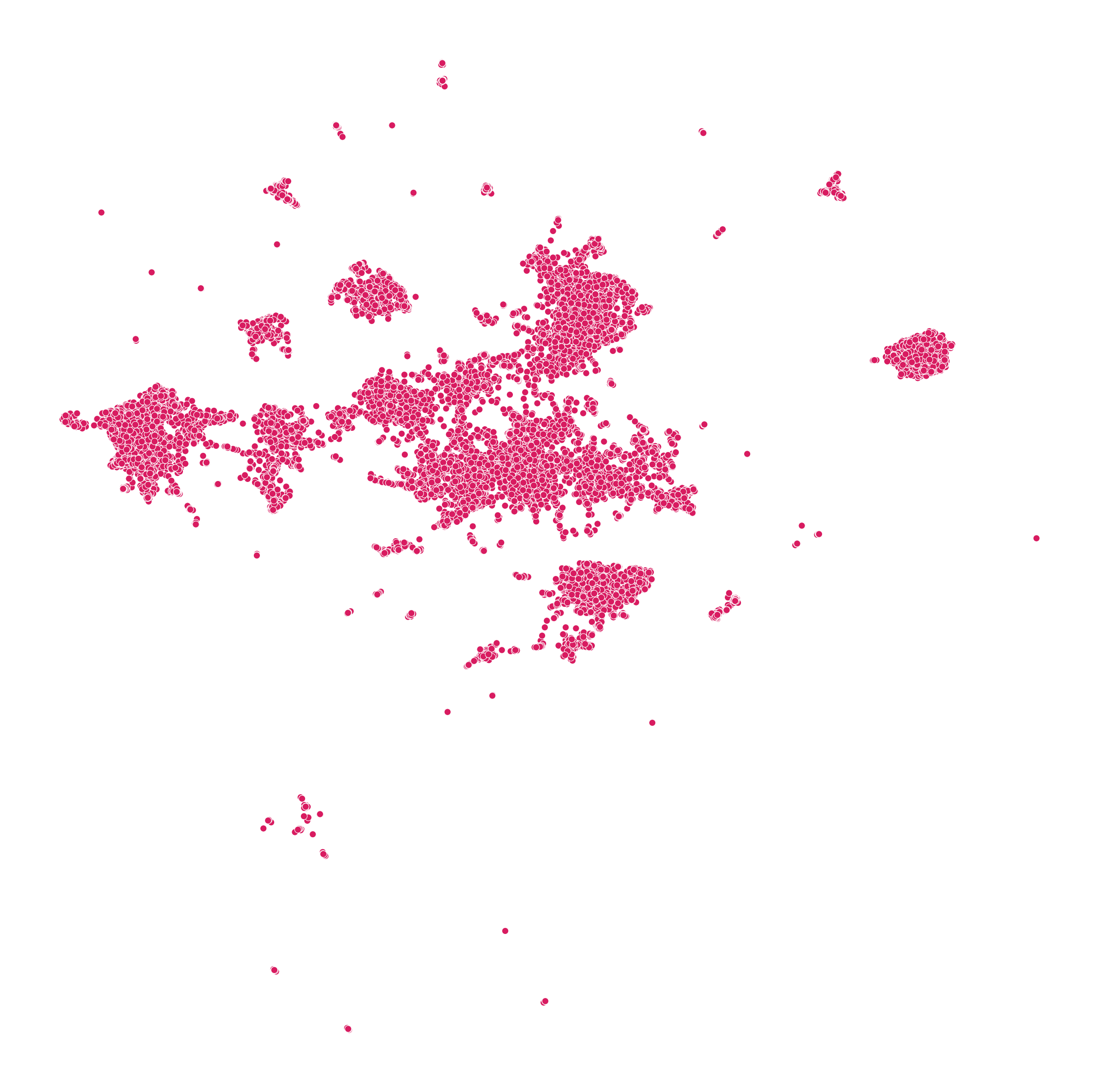}
        \caption{\experiment{2C}}
    \end{subfigure}
    \caption{UMAP projections of intent contribution explanations across experiments \experiment{2A}, \experiment{2B} \experiment{2C}. Each point represents one tweet-model pair; color encodes ground truth (blue: non-misleading, red: misleading). The progressive dispersion of the embedding space reflects the increasing prior knowledge provided to the model across conditions.}
    \label{fig:umap_intent}
\end{figure}

\para{Semantic Analysis}
UMAP projections of intent explanations (\cref{fig:umap_intent}) replicate the progressive dispersion pattern observed for rhetoric. \experiment{2A} shows a more compressed semantic space than the corresponding rhetoric condition (mean nearest-neighbor distance $0.0012$ vs.\ $0.0015$, silhouette $0.007$ vs.\ $0.015$), consistent with the greater inferential difficulty of attributing latent authorial intent.

\para{Summary}
Intent attribution is inferentially harder than rhetoric identification: semantic spaces are more compressed, ESS gains from condition \rcond{C} are smaller, and the negative scale-ESS correlation persists even with error anchoring. \kimi\ is the only model exceeding the human ESS reference. No model replicates the full expert cross-family intent structure; \gta\ comes closest for Aesthetic-Driven Misrepresentation and \gpt\ for Bias Exploitation.

\section{VisLies Gallery Analysis}
\label{sec:vislies}

To assess the main findings beyond the COVID-19 domain, we applied the same experimental study on the VisLies gallery~\cite{vislies}, a community-curated collection of real-world examples of visualizations affected by perceptual, cognitive, and conceptual errors. VisLies is an IEEE VIS community event dedicated to showcasing deceptive and misleading visualizations.
From this gallery, we collected 130 visualizations, all of which are misleading, spread over different types and domains. Unlike the Twitter dataset, we have only misleading examples, and there is no ground-truth error mapping to the error taxonomy used for that dataset.
This limits the analysis to conditions \rcond{A} and \rcond{B} in their original form, while condition \rcond{C} can still be applied by supplying each model with the VisLies item descriptions curated by the community as a proxy for broader ground-truth error information.
We collected and structured these 130 visualizations in a computer-readable format as a reliable additional test since all of them are truly misleading and were identified through collective community-expert analysis.
We plan to release this dataset after contacting the organizers of VisLies for future reference and use.

\para{RQ0 – Identify Misleading Visualizations}
Since the dataset is entirely positive, evaluating this research question in \experiment{1A} and \experiment{2A} involves focusing on recall (true positive rate).
Results are reported in \cref{fig:vislies_recall}.
\gemma\ achieves the highest recall in both conditions ($0.992$), followed by \kimi ($0.946$/$0.923$) and \gpt ($0.923$/$0.938$).
However, this result must be interpreted with caution for \gemma: it was previously identified as a strongly positively-biased affected classifier that assigns misleading labels indiscriminately.
The near-perfect recall on the VisLies dataset is therefore an effect of that bias, and is fully consistent with the low MCC values it obtained on the balanced COVID dataset.
By contrast, \gpt, which achieved strong MCC on COVID, here produces a comparably high recall through correct detection rather than indiscriminate labeling, and can be considered the only top-tier model for which the recall is informative.
Notably, the bottom of both the rankings includes \intern, which performed well in the COVID dataset but failed here (recall $\approx 0.3$).
Models that have already been known to have low MCC (e.g., \llava, \deepseek) continue to show low recall here, confirming that their low MCC reflects intrinsic detection weakness rather than conservatism.
The rank stability between conditions shows notable shuffling in the middle tier, indicating that task framing continues to influence detection even on unambiguous stimuli.

\para{RQ1 -- Identify Rhetorical Techniques}
The dominant rhetoric assigned by the models in the three experimental conditions (\cref{fig:rhet_profile_vislies}) is \textit{Mapping}, followed consistently by \textit{Information} and \textit{Linguistic}. 
This order constitutes the most informative contrast with the COVID dataset results, in which Information Access emerged as the primary failure mode across conditions.
On VisLies, this order is reversed.
The inversion is expected: VisLies is a corpus of classic deceptions, heavily populated by design-level distortions such as truncated axes, area, and 3D encodings.
These are precisely the error types that should be associated with elevated \textit{Mapping} rhetoric scores.
The COVID dataset, by contrast, contains a substantial proportion of reasoning errors (e.g., cherry-picking, causal inference) that manifest primarily through \textit{Information Access} rhetoric.
The cross-dataset comparison, therefore, provides independent empirical support for the error-rhetoric coupling: \textit{Mapping} rhetoric tracks design violations, and \textit{Information Access} tracks reasoning errors, with the predominant error type in each dataset predicting the observed rhetoric hierarchy.

\para{RQ2 -- Identify Authorial Intents}
The intent profiles across all three conditions (\cref{fig:intent_profile_vislies}) show \textit{Context Distortion} as the dominant category, consistent across models and conditions. \textit{Lack of visualization literacy} increases in \experiment{2C}, since the additional information provided (a description taken from the VisLies gallery) includes precise details about deceptions.
This predominance is consistent with the nature of the collection, which typically decontextualizes data or frames it to highlight its misleadingness, rather than embedding it within rhetorical narrative constructions characteristic of COVID-19 social media misinformation.
Moving from \rcond{A} to \rcond{B} highly increases the scores assigned to intentional deception intents, with \kimi, \gpt, and \qwen\ with the highest overall rating. This result indicates that being told the visualization is misleading increases attribution weight across all intent categories: weaker models tend to increase scores evenly, while larger models tend to increase scores especially for intentional ones.
This is consistent with the finding that \gpt\ and \maverick\ produce the most structured and diagnostically informative attribution profiles under the anchored explanation conditions.

\section{Limitations and Conclusion}
\label{sec:discussion}

Several limitations should be noted. 
The UMAP projections and the preliminary BERTopic analyses presented in this work should be understood as a form of \textit{latent space} analysis. Structural features of the embedding space, such as class separability and cluster topology, could reveal how models organize their reasoning.
Future work could explore this space more deeply.
To support an initial exploration, we provide a web-based tool as supplemental material.
We evaluated the largest variant of each model family; however, the field is increasingly oriented toward small, efficient models for edge deployment (SLMs) \cite{Lu2025,JiachengWang2025,RuiWang2025}, and it remains an open question whether smaller variants within the same family would exhibit comparable attribution behavior, especially given the negative correlation between model parameters and ESS. 
The COVID-19 dataset uses 14 error types from Lisnic et al. \cite{Lisnic2023}; annotating both COVID-19 and VisLies against the finer-grained 74-type taxonomy of Lo et al. \cite{LoGupta2022} (\cref{sec:errors_def}) would enable more granular analysis. Similarly, mapping VisLies to the Lisnic et al. error types would allow a full condition \rcond{C} replication on an independent corpus. 
The human baseline was collected from visualization experts; involving less-experienced users could provide a different baseline for comparison.
Each experiment used a single prompt per condition, and we tested the default temperature for each model, so our rankings reflect a specific phrasing~\cite{Sclar2024}. 

In conclusion, this paper evaluated LLMs on their ability to detect misleading visualizations, recognize rhetorical techniques, and attribute authorial intent, using visualizations from COVID-19 tweets and the VisLies gallery across six experiments with increasing levels of prior knowledge. We plan to address the reported limitations as future work.
The datasets, all the generated data from the tested LLMs, and a web-based explorer of results are available at {\small\url{https://github.com/XAIber-lab/truevislies}}.

\bibliographystyle{abbrv-doi-hyperref}
\bibliography{biblio}

\clearpage
\newpage
\appendix

\section{Appendix}
The appendix reports visual support materials to the analyses discussed in the paper for cross-reference and legibility. They cover the visual interactive explorer, the LLMs characteristics, the benchmarking results on the three research questions for the COVID-19 dataset, and for the VisLies dataset.

\subsection{Interactive Visual Explorer}
\begin{figure}[h]
    \centering
    \includegraphics[width=\linewidth]{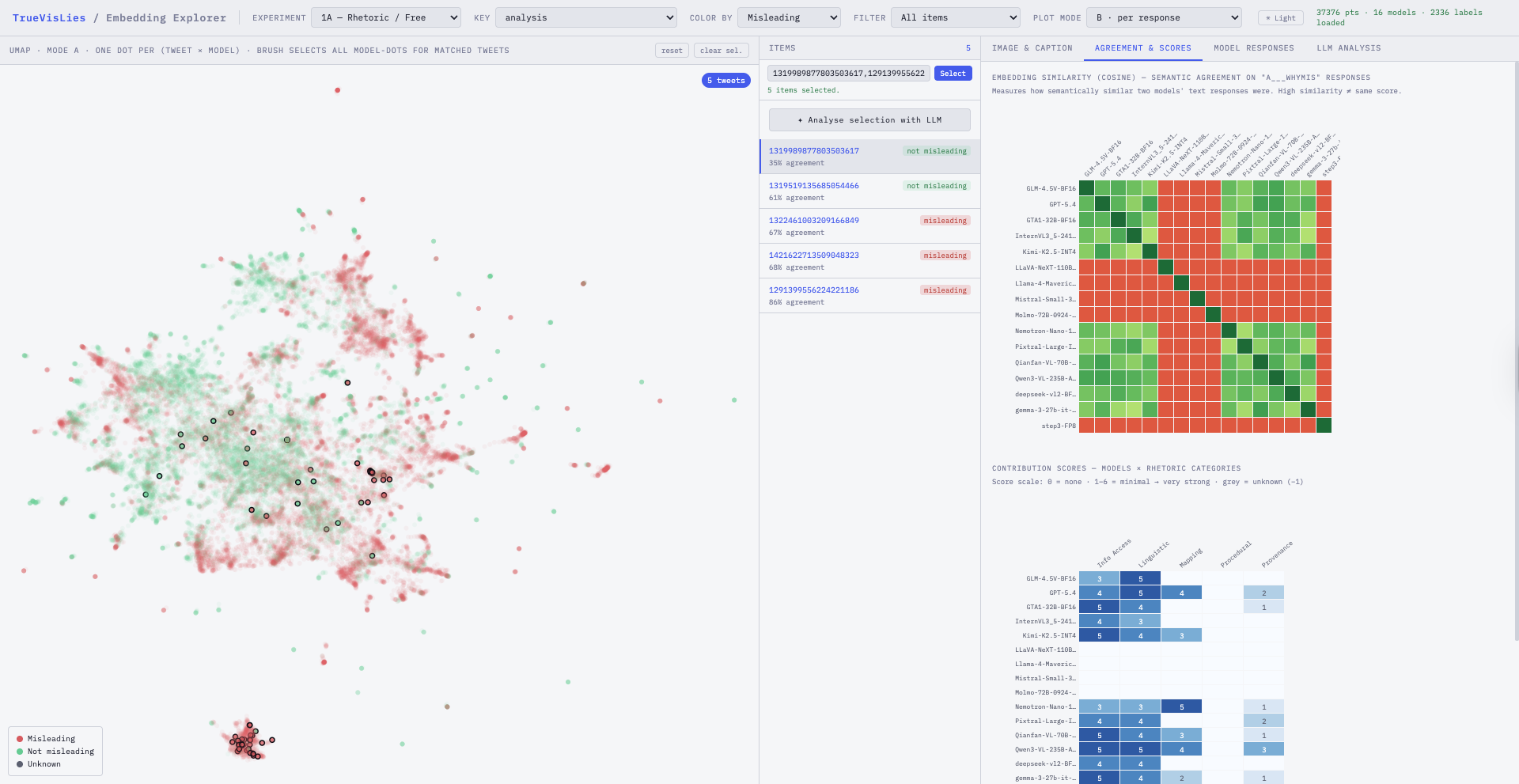}
    \caption{Explorer showing the images highlighted in the teaser.}
    \label{fig:explorer}
\end{figure}

Figure~\ref{fig:explorer} reports the implemented visual explorer of the results discussed in the paper, configured on the \experiment(1A) visible in the paper teaser image.

\begin{table*}[t]
\centering
\caption{Detailed overview of the 15 open-source LLMs evaluated in this study.
Parameters are reported in billions (B).
For Mixture-of-Experts (MoE) models, the \textit{Active} column reports the number of active parameters.
The \textit{ID} column provides the short nickname used throughout the paper and figures.
Models are grouped by total parameter count into three groups: Small ($\leq$33\,B), Medium (70--124\,B), and Large ($\geq$235\,B).
All models were executed using their official vendor-provided weights and default precision. We used the vendor-provided reduced-precision variants (FP8) for \maverick\ and \stepthree\ because their default precision exceeds our available GPU memory.
}
\label{tab:models-detail}
\small
\setlength{\tabcolsep}{4.5pt}
\begin{tabular}{@{}lllcrrlc@{}}
\toprule
\textbf{ID} & \textbf{Model} & \textbf{Provider} & \textbf{Architecture} & \multicolumn{2}{c}{\textbf{Params (B)}} & \textbf{Hugging Face Model ID} & \\
\cmidrule(lr){5-6}
& & & & \textbf{Total} & \textbf{Active} & & \\

\midrule
\multicolumn{8}{@{}l}{\textit{Small models (Total $\leq$ 33\,B)}} \\[10pt]

\nemotron & Nemotron-Nano-V2-VL & Nvidia \cite{nvidia2025nvidianemotronnanov2} & Dense & 12 & -- & \texttt{nvidia/NVIDIA-Nemotron-Nano-12B-v2-VL-BF16} & \hflink{nvidia/NVIDIA-Nemotron-Nano-12B-v2-VL-BF16} \\

\mistral & Mistral-Small-3.2 & Mistral AI & Dense & 24 & -- & \texttt{mistralai/Mistral-Small-3.2-24B-Instruct-2506} & \hflink{mistralai/Mistral-Small-3.2-24B-Instruct-2506} \\

\deepseek & DeepSeek-VL2 & DeepSeek \cite{wu2024deepseekvl2} & MoE & 27 & 5 & \texttt{deepseek-ai/deepseek-vl2} & \hflink{deepseek-ai/deepseek-vl2} \\

\gemma & Gemma3 & Google \cite{gemmateam2025gemma3technicalreport} & Dense & 27 & -- & \texttt{google/gemma-3-27b-it} & \hflink{google/gemma-3-27b-it} \\

\gta & GTA1 & Salesforce \cite{yang2025gta1guitesttimescaling} & Dense & 32 & -- & \texttt{Salesforce/GTA1-32B} & \hflink{Salesforce/GTA1-32B} \\

\midrule
\multicolumn{8}{@{}l}{\textit{Medium models (70--124\,B total)}} \\[10pt]

\qianfan & Qianfan-VL & Baidu \cite{qianfan-vl-2025} & Dense & 70 & -- & \texttt{baidu/Qianfan-VL-70B} & \hflink{baidu/Qianfan-VL-70B} \\

\molmo & Molmo & Ai2 \cite{deitke2024molmopixmoopenweights} & Dense & 72 & -- & \texttt{allenai/Molmo-72B-0924} & \hflink{allenai/Molmo-72B-0924} \\

\glm & GLM-4.5V & Z.ai \cite{vteam2025glm45} & MoE & 108 & 12 & \texttt{zai-org/GLM-4.5V} & \hflink{zai-org/GLM-4.5V} \\

\llava & LLaVA-NeXT & LLaVA \cite{li2024llavanext} & Dense & 110 & -- & \texttt{llava-hf/llava-next-110b-hf} & \hflink{llava-hf/llava-next-110b-hf} \\

\pixtral & Pixtral-Large & Mistral AI & Dense & 124 & -- & \texttt{mistralai/Pixtral-Large-Instruct-2411} & \hflink{mistralai/Pixtral-Large-Instruct-2411} \\

\midrule
\multicolumn{8}{@{}l}{\textit{Large models (Total $\geq$ 235\,B)}} \\[10pt]

\qwen & Qwen3-VL & Alibaba \cite{qwen3technicalreport} & MoE & 235 & 22 & \texttt{Qwen/Qwen3-VL-235B-A22B-Instruct} & \hflink{Qwen/Qwen3-VL-235B-A22B-Instruct} \\

\intern & InternVL3.5 & OpenGVLab \cite{wang2025internvl} & MoE & 241 & 28 & \texttt{OpenGVLab/InternVL3\_5-241B-A28B} & \hflink{OpenGVLab/InternVL3_5-241B-A28B} \\

\stepthree & Step3 & StepFun AI \cite{step3system} & MoE & 321 & 38 & \texttt{stepfun-ai/step3-fp8} & \hflink{stepfun-ai/step3-fp8} \\

\maverick & Llama-4-Maverick & Meta & MoE & 400 & 17 & \texttt{meta-llama/Llama-4-Maverick-17B-128E-Instruct-FP8} & \hflink{meta-llama/Llama-4-Maverick-17B-128E-Instruct-FP8} \\

\kimi & Kimi-K2.5 & Moonshot AI \cite{kimiteam2026kimik25} & MoE & 1,000 & 32 & \texttt{moonshotai/Kimi-K2.5} & \hflink{moonshotai/Kimi-K2.5} \\

\midrule
\gpt & GPT-5.4-2026-03-05 & OpenAI & -- & -- & -- & -- \\

\bottomrule
\end{tabular}
\end{table*}

\subsection{LLMs characteristics}

\clearpage
\newpage

\begin{figure*}[t]
\centering
\includegraphics[width=0.7\linewidth]{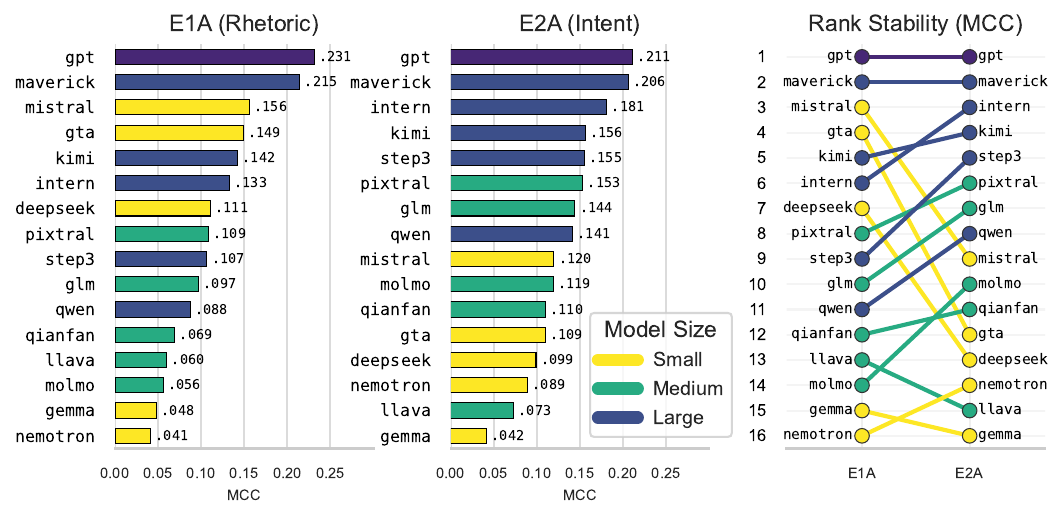}
\caption{
MCC scores for all 16 models in \experiment{1A} (Rhetoric) and \experiment{2A} (Intent), with rank stability between the two conditions shown on the right. Overall performance is low in both experiments (mean MCC $0.113$ and $0.132$, respectively), with \gpt\ and \maverick\ leading consistently. 
Color encodes model size (small, medium, large); larger models tend to score higher, a trend that is statistically significant in \experiment{2A}.
}
\label{fig:app-mcc}
\end{figure*}

\begin{figure*}[t]
\centering
\includegraphics[width=0.7\linewidth]{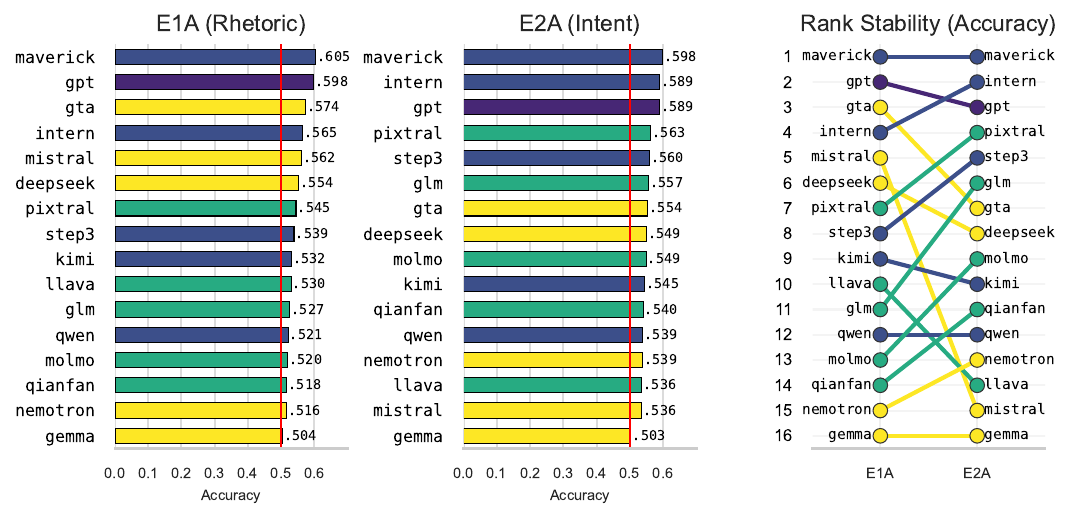}
\caption{
Accuracy scores for all 16 models in \experiment{1A} (Rhetoric) and \experiment{2A} (Intent), with rank stability between the two conditions shown on the right.
Overall performance is low in both experiments (mean accuracy $0.544$ and $0.553$, respectively), with \maverick\ leading consistently. 
Color encodes model size (small, medium, large); larger models tend to score higher.
}
\label{fig:app-accuracy}
\end{figure*}

\begin{table*}[t]
\caption{Summary of classification results for experiment \experiment{1A}}
\label{tab:classification-summary-1A}
\centering
\small
\begin{tabular}{l|rr|rr|rr|rr|rr}
\toprule

\textbf{Model Size} &
\multicolumn{2}{c|}{\textbf{MCC}} &
\multicolumn{2}{c|}{\textbf{Accuracy}} &
\multicolumn{2}{c|}{\textbf{F1}} &
\multicolumn{2}{c|}{\textbf{Precision}} &
\multicolumn{2}{c}{\textbf{Recall}} \\
\cmidrule(lr){2-3}
\cmidrule(lr){4-5}
\cmidrule(lr){6-7}
\cmidrule(lr){8-9}
\cmidrule(lr){10-11}
& $\mu$ & $\sigma$ & $\mu$ & $\sigma$ & $\mu$ & $\sigma$
& $\mu$ & $\sigma$ & $\mu$ & $\sigma$ \\
\midrule

Small & 0.101 & 0.054 & 0.542 & 0.030 & 0.622 & 0.052 & 0.536 & 0.033 & 0.775 & 0.189 \\
Medium & 0.078 & 0.023 & 0.528 & 0.011 & 0.632 & 0.050 & 0.519 & 0.009 & 0.830 & 0.159 \\
Large & \textbf{0.153} & 0.058 & \textbf{0.560} & 0.035 & \textbf{0.640} & 0.050 & 0.551 & 0.046 & 0.805 & 0.190 \\

\midrule
ALL & 0.113 & 0.056 & 0.544 & 0.030 & 0.632 & 0.048 & 0.536 & 0.035 & 0.804 & 0.170 \\

\bottomrule
\end{tabular}
\end{table*}
\begin{table*}[t]
\caption{Summary of classification results for experiment \experiment{2A}}
\label{tab:classification-summary-2A}
\centering
\small
\begin{tabular}{l|rr|rr|rr|rr|rr}
\toprule

\textbf{Model Size} &
\multicolumn{2}{c|}{\textbf{MCC}} &
\multicolumn{2}{c|}{\textbf{Accuracy}} &
\multicolumn{2}{c|}{\textbf{F1}} &
\multicolumn{2}{c|}{\textbf{Precision}} &
\multicolumn{2}{c}{\textbf{Recall}} \\
\cmidrule(lr){2-3}
\cmidrule(lr){4-5}
\cmidrule(lr){6-7}
\cmidrule(lr){8-9}
\cmidrule(lr){10-11}
& $\mu$ & $\sigma$ & $\mu$ & $\sigma$ & $\mu$ & $\sigma$
& $\mu$ & $\sigma$ & $\mu$ & $\sigma$ \\
\midrule

Small & 0.092 & 0.030 & 0.536 & 0.020 & 0.612 & 0.062 & 0.531 & 0.023 & 0.761 & 0.214 \\
Medium & 0.120 & 0.032 & 0.549 & 0.011 & \textbf{0.634} & 0.050 & 0.533 & 0.006 & 0.796 & 0.137 \\
Large & \textbf{0.175} & 0.029 & \textbf{0.570} & 0.025 & 0.630 & 0.068 & 0.564 & 0.048 & 0.769 & 0.226 \\

\midrule
ALL & 0.132 & 0.046 & 0.553 & 0.024 & 0.626 & 0.058 & 0.544 & 0.034 & 0.775 & 0.186 \\

\bottomrule
\end{tabular}
\end{table*}

\begin{table*}[t]
\caption{Detailed classification results for experiment \experiment{1A}, Model are sorted by MCC. Best values for each metric are reported in bold.}
\label{tab:classification-details-1A}
\centering
\small
\begin{tabular}{rlrrrrrrrrr}
\toprule
\textbf{Rank} & \textbf{Model} & \textbf{MCC} & \textbf{Accuracy} & \textbf{F1} & \textbf{Precision} & \textbf{Recall} & \textbf{TN} & \textbf{FP} & \textbf{FN} & \textbf{TP} \\
\midrule

1 & gpt & \textbf{0.231} & 0.598 & \textbf{0.682} & 0.564 & 0.863 & 389 & 779 & 160 & 1008 \\
2 & maverick & 0.215 & \textbf{0.605} & 0.558 & 0.633 & 0.499 & 830 & 338 & 585 & 583 \\
3 & mistral & 0.156 & 0.562 & 0.664 & 0.538 & 0.868 & 298 & 870 & 154 & 1014 \\
4 & gta & 0.149 & 0.574 & 0.542 & 0.585 & 0.504 & 751 & 417 & 579 & 589 \\
5 & kimi & 0.142 & 0.532 & 0.677 & 0.517 & 0.979 & 98 & 1070 & 24 & 1144 \\
6 & intern & 0.133 & 0.565 & 0.599 & 0.556 & 0.650 & 562 & 606 & 409 & 759 \\
7 & deepseek & 0.111 & 0.554 & 0.604 & 0.543 & 0.682 & 496 & 672 & 371 & 797 \\
8 & pixtral & 0.109 & 0.545 & 0.645 & 0.529 & 0.825 & 309 & 859 & 204 & 964 \\
9 & step3 & 0.107 & 0.539 & 0.656 & 0.523 & 0.881 & 230 & 938 & 139 & 1029 \\
10 & glm & 0.097 & 0.527 & 0.666 & 0.514 & 0.945 & 126 & 1042 & 64 & 1104 \\
11 & qwen & 0.088 & 0.521 & 0.667 & 0.511 & 0.960 & 96 & 1072 & 47 & 1121 \\
12 & qianfan & 0.069 & 0.518 & 0.663 & 0.509 & 0.949 & 101 & 1067 & 60 & 1108 \\
13 & llava & 0.060 & 0.530 & 0.544 & 0.528 & 0.561 & 583 & 585 & 513 & 655 \\
14 & molmo & 0.056 & 0.520 & 0.644 & 0.512 & 0.868 & 201 & 967 & 154 & 1014 \\
15 & gemma & 0.048 & 0.504 & 0.668 & 0.502 & 0.997 & 12 & 1156 & 3 & 1165 \\
16 & nemotron & 0.041 & 0.516 & 0.630 & 0.510 & 0.824 & 243 & 925 & 205 & 963 \\

\bottomrule
\end{tabular}
\end{table*}
\begin{table*}[t]
\caption{Detailed classification results for experiment \experiment{2A}, Model are sorted by MCC. Best values for each metric are reported in bold.}
\label{tab:classification-details-2A}
\centering
\small
\begin{tabular}{rlrrrrrrrrr}
\toprule
\textbf{Rank} & \textbf{Model} & \textbf{MCC} & \textbf{Accuracy} & \textbf{F1} & \textbf{Precision} & \textbf{Recall} & \textbf{TN} & \textbf{FP} & \textbf{FN} & \textbf{TP} \\

\midrule

1 & gpt & \textbf{0.211} & 0.589 & \textbf{0.677} & 0.557 & 0.860 & 370 & 798 & 163 & 1005 \\
2 & maverick & 0.206 & \textbf{0.598} & 0.530 & 0.639 & 0.452 & 870 & 298 & 640 & 528 \\
3 & intern & 0.181 & 0.589 & 0.555 & 0.606 & 0.513 & 778 & 390 & 569 & 599 \\
4 & kimi & 0.156 & 0.545 & \textbf{0.677} & 0.525 & 0.954 & 159 & 1009 & 54 & 1114 \\
5 & step3 & 0.155 & 0.560 & 0.666 & 0.536 & 0.880 & 279 & 889 & 140 & 1028 \\
6 & pixtral & 0.153 & 0.563 & 0.659 & 0.541 & 0.844 & 330 & 838 & 182 & 986 \\
7 & glm & 0.144 & 0.557 & 0.661 & 0.535 & 0.863 & 293 & 875 & 160 & 1008 \\
8 & qwen & 0.141 & 0.539 & 0.675 & 0.522 & 0.955 & 145 & 1023 & 53 & 1115 \\
9 & mistral & 0.120 & 0.536 & 0.668 & 0.520 & 0.933 & 163 & 1005 & 78 & 1090 \\
10 & molmo & 0.119 & 0.549 & 0.649 & 0.531 & 0.835 & 307 & 861 & 193 & 975 \\
11 & qianfan & 0.110 & 0.540 & 0.658 & 0.524 & 0.884 & 228 & 940 & 135 & 1033 \\
12 & gta & 0.109 & 0.554 & 0.527 & 0.562 & 0.496 & 716 & 452 & 589 & 579 \\
13 & deepseek & 0.099 & 0.549 & 0.570 & 0.545 & 0.598 & 585 & 583 & 470 & 698 \\
14 & nemotron & 0.089 & 0.539 & 0.629 & 0.526 & 0.783 & 345 & 823 & 254 & 914 \\
15 & llava & 0.073 & 0.536 & 0.544 & 0.535 & 0.554 & 606 & 562 & 521 & 647 \\
16 & gemma & 0.042 & 0.503 & 0.667 & 0.502 & 0.997 & 12 & 1156 & 4 & 1164 \\

\bottomrule
\end{tabular}
\end{table*}

\begin{figure*}[t]
    \centering
    \begin{subfigure}{0.35\linewidth}
        \centering
        \includegraphics[width=\linewidth]{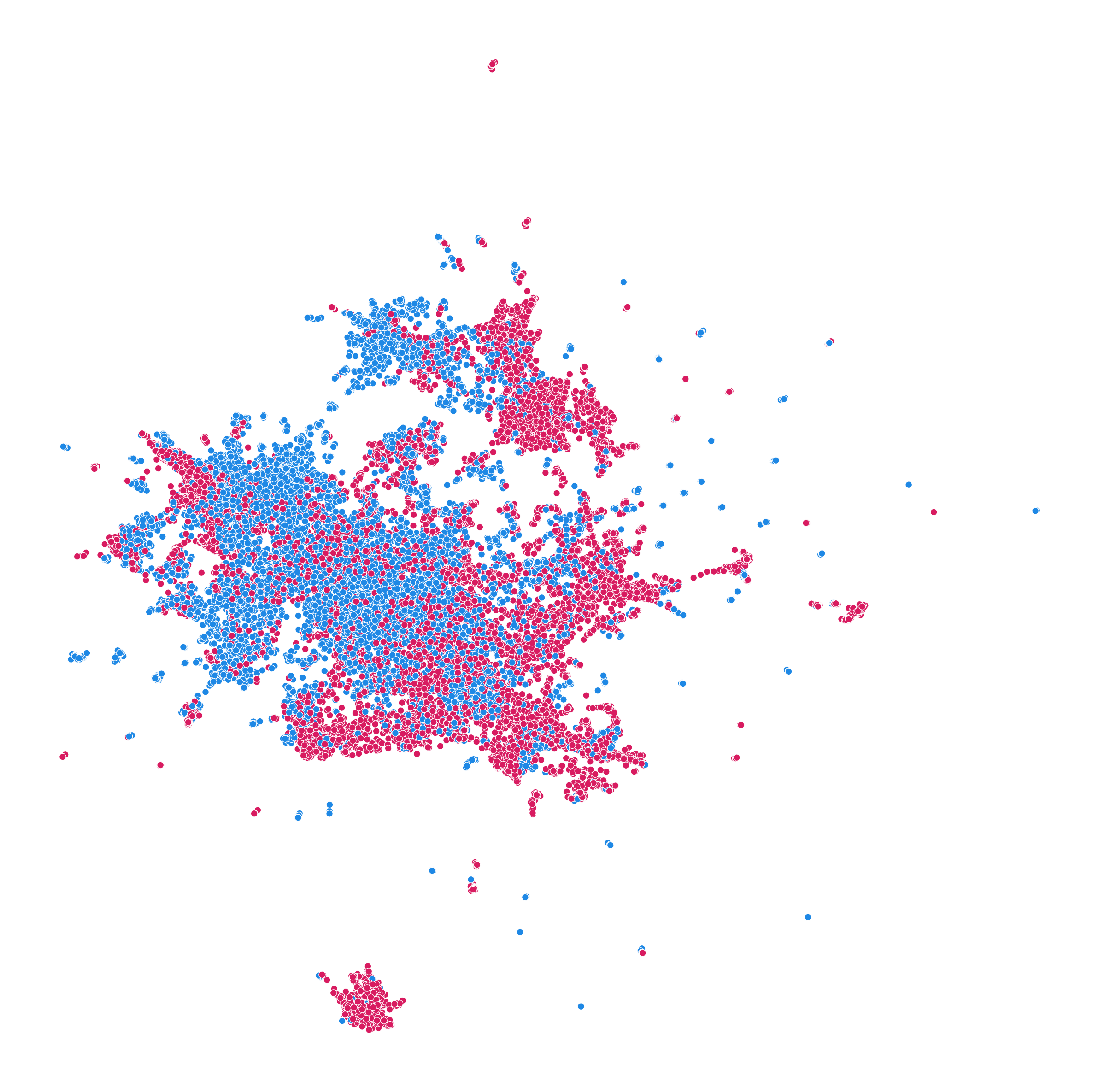}
        \caption{\experiment{1A} - Rhetoric}
        \label{fig:a}
    \end{subfigure}
    \begin{subfigure}{0.35\linewidth}
        \centering
        \includegraphics[width=\linewidth]{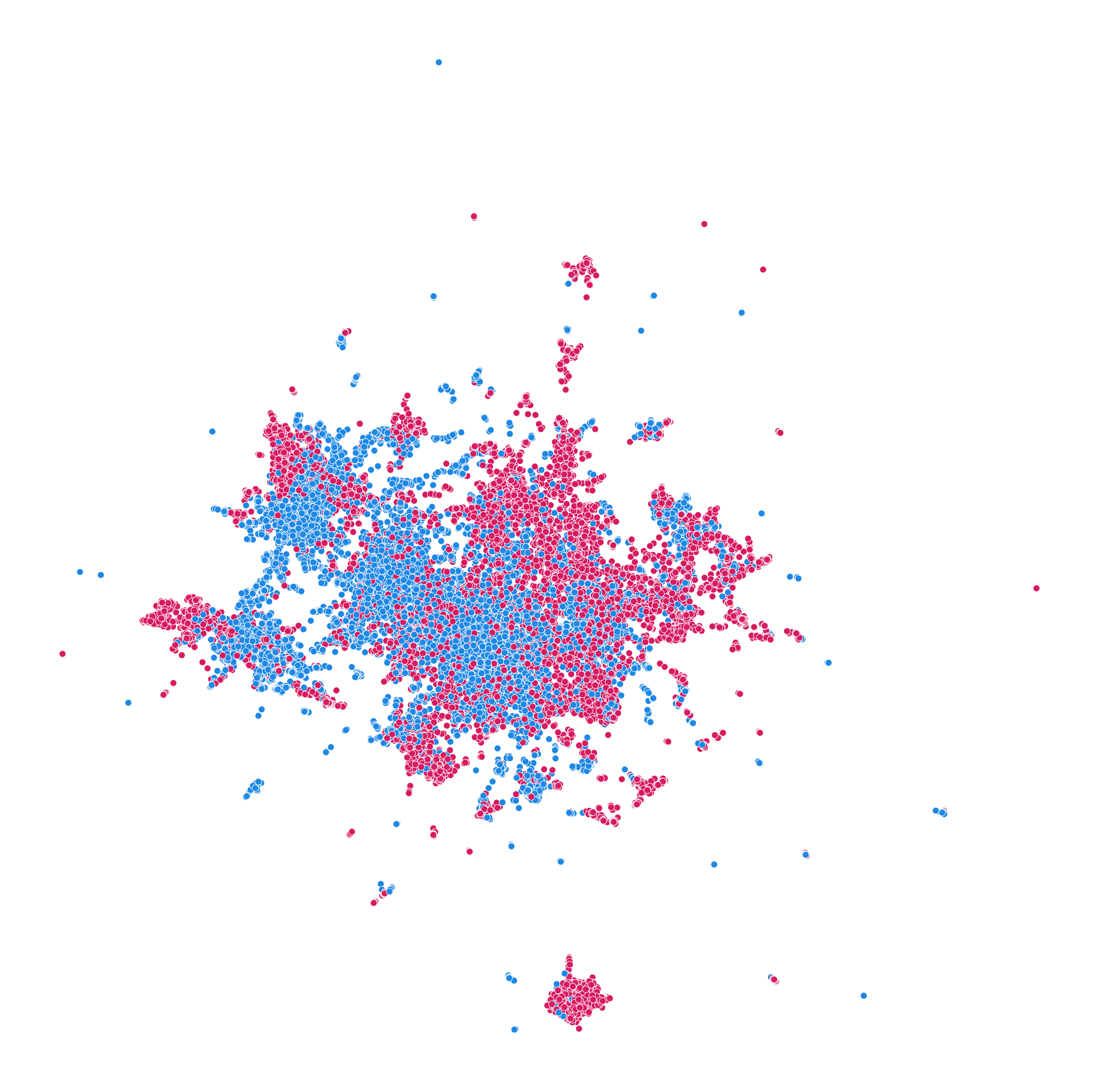}
        \caption{\experiment{2A} - Intent}
        \label{fig:b}
    \end{subfigure}
    \caption{UMAP projections of model explanations for why a visualization is perceived as misleading, for \experiment{1A} (rhetoric analysis) and \experiment{2A} (intent analysis). Each point represents one tweet-model pair; color encodes ground truth (blue: non-misleading, red: misleading). Both projections form a single, densely packed cloud with no visible class boundary and small mean nearest-neighbor distances, indicating that the models produce semantically similar explanations regardless of the ground truth. The spread of truly misleading explanations (red) is larger than that of non-misleading ones (blue), reflecting the diversity of error types present in the dataset.}
    \label{fig:umap_whymis}
\end{figure*}


\begin{figure*}[t]
    \centering
    \begin{subfigure}[t]{0.7\linewidth}
        \centering
        \includegraphics[width=\linewidth]{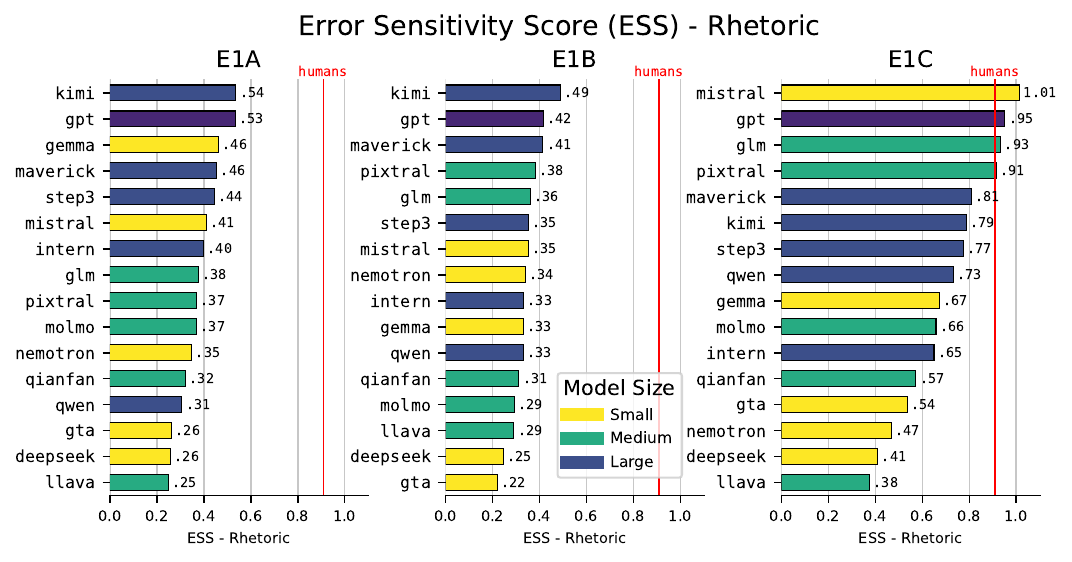}
        \caption{Rhetoric (\experiment{1A}, \experiment{1B}, \experiment{1C})}
        \label{fig:ess_rhetoric}
    \end{subfigure}
    \begin{subfigure}[t]{0.7\linewidth}
        \centering
        \includegraphics[width=\linewidth]{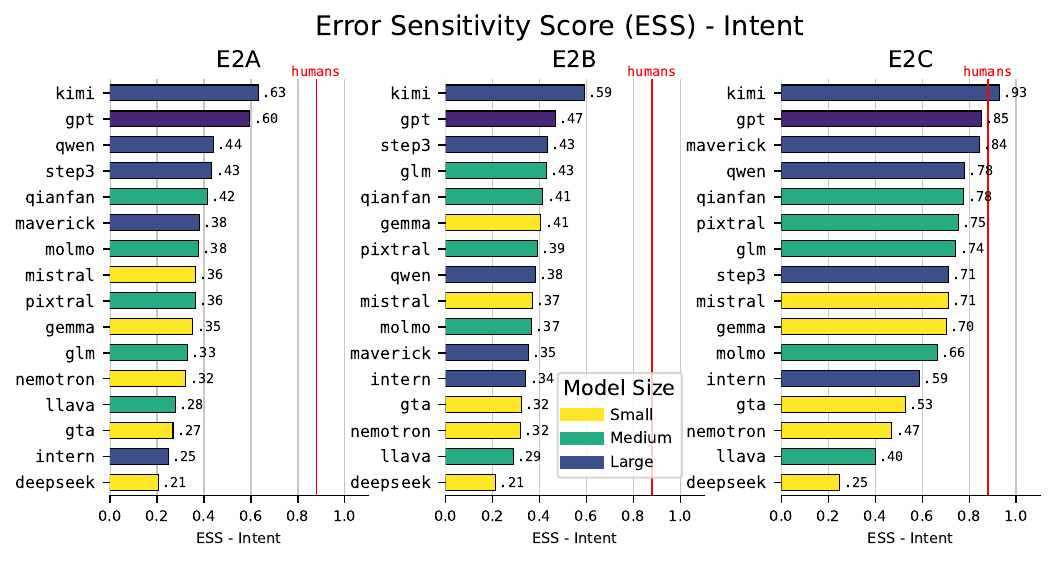}
        \caption{Intent (\experiment{2A}, \experiment{2B}, \experiment{2C})}
        \label{fig:ess_intent}
    \end{subfigure}
    \caption{Error Sensitivity Score (ESS) per model and experimental condition for rhetoric (a) and intent (b). Higher ESS indicates sharper discrimination across error types. Models are sorted by ESS; the human expert reference is shown as a red line. Color encodes model size group.}
    \label{fig:ess}
\end{figure*}

\clearpage

\begin{figure*}[t]
    \centering
    \begin{subfigure}[t]{0.32\linewidth}
        \centering
        \includegraphics[width=\linewidth]{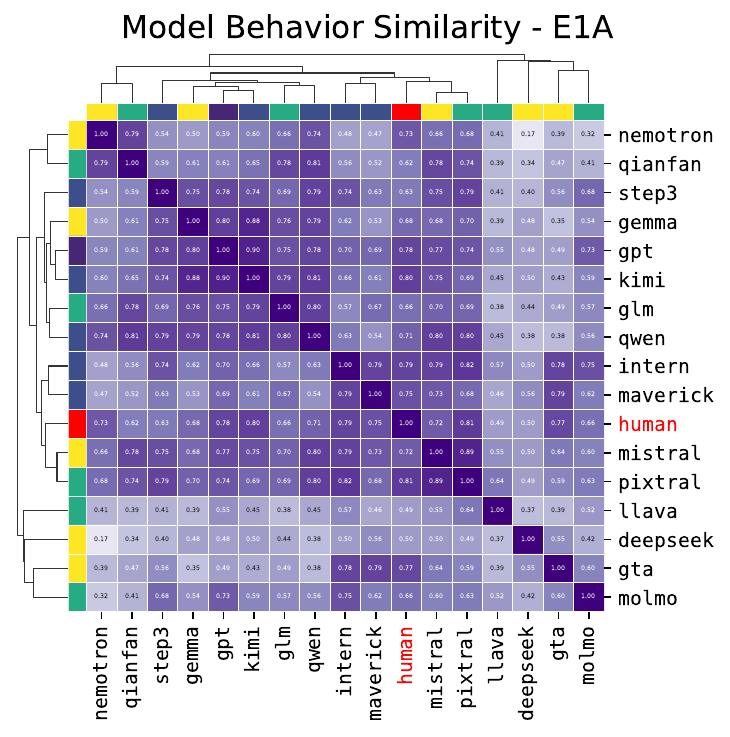}
        \caption{\experiment{1A}}
        \label{fig:similarity_rhetoric_E1A}
    \end{subfigure}
    \hfill
    \begin{subfigure}[t]{0.32\linewidth}
        \centering
        \includegraphics[width=\linewidth]{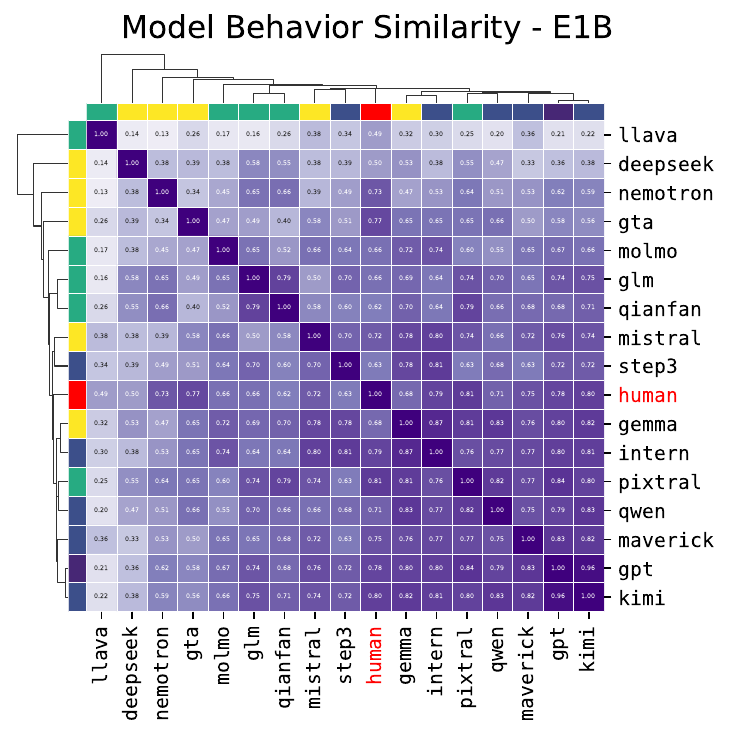}
        \caption{\experiment{1B}}
        \label{fig:similarity_rhetoric_E1B}
    \end{subfigure}
    \hfill
    \begin{subfigure}[t]{0.32\linewidth}
        \centering
        \includegraphics[width=\linewidth]{figures/similarity_matrix_E1C.pdf}
        \caption{\experiment{1C}}
        \label{fig:similarity_rhetoric_E1C}
    \end{subfigure}
    \caption{%
        Pairwise behavioral similarity matrices for rhetoric across the three experimental conditions.
        Rows and columns are ordered by hierarchical clustering.
        The similarity range compresses progressively from \experiment{1A} to \experiment{1C}, reflecting behavioral convergence induced by the ground-truth error label.
        \llava\ and \deepseek\ are the most isolated models across all
        conditions; \model{gpt} and \model{kimi} form the most stable high-similarity pair.
        Human similarity with the model population remains moderate even in \experiment{1C}.
    }
    \label{fig:similarity_rhetoric_app}
\end{figure*}


\begin{figure*}[t]
    \centering
    \begin{subfigure}[t]{0.3\linewidth}
        \centering
        \includegraphics[width=\linewidth]{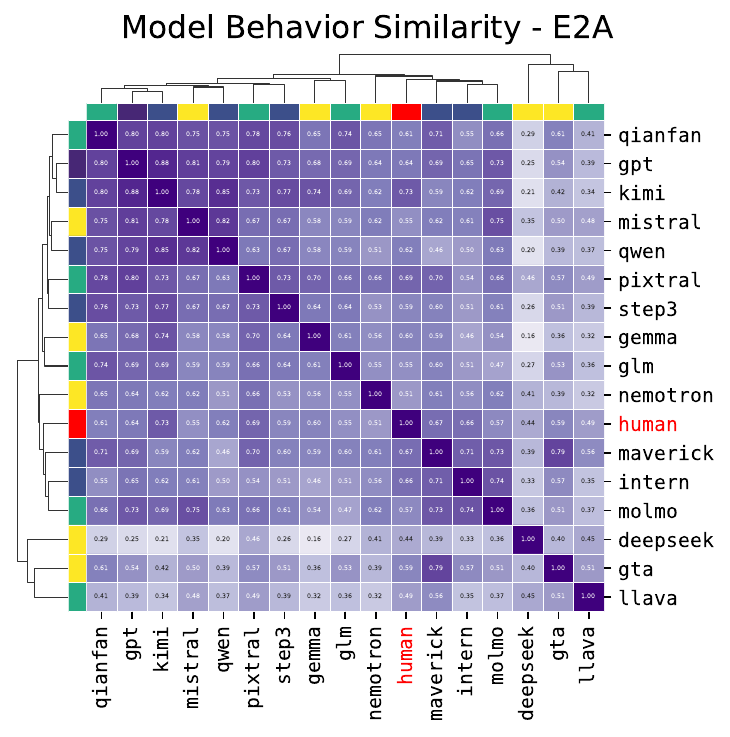}
        \caption{\experiment{E2A}}
        \label{fig:sim_intent_a}
    \end{subfigure}
    \begin{subfigure}[t]{0.3\linewidth}
        \centering
        \includegraphics[width=\linewidth]{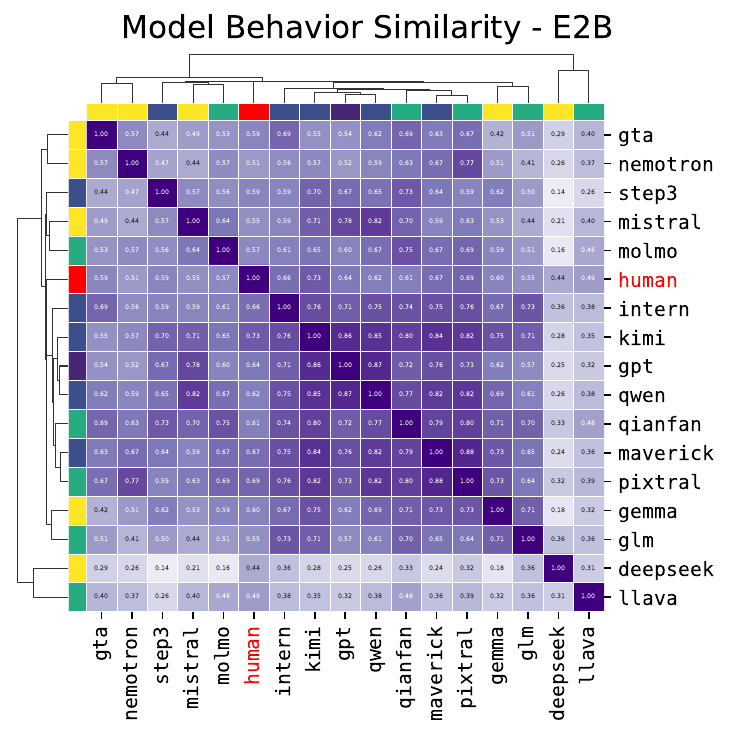}
        \caption{\experiment{E2B}}
        \label{fig:sim_intent_b}
    \end{subfigure}
     \begin{subfigure}[t]{0.3\linewidth}
        \centering
        \includegraphics[width=\linewidth]{figures/similarity_matrix_E2C.pdf}
        \caption{\experiment{E2C}}
        \label{fig:sim_intent_c}
    \end{subfigure}
    \caption{Pairwise behavioral similarity matrices for authorial intent attribution across the three experimental conditions.
    Rows and columns are ordered by hierarchical clustering.
    The similarity range compresses progressively from \experiment{1A} to \experiment{1C}, reflecting behavioral convergence induced by the ground-truth error label.
    \deepseek\ is pronounced outlier.
    By \rcond{C}, the minimum pairwise similarity rises and a dominant convergence cluster emerges.
    Compared with the rhetoric similarity matrices, the intent matrices exhibit greater spread across all conditions.}
    \label{fig:similarity_intent_app}
\end{figure*}


\clearpage
\newpage

\begin{figure*}[h]
    \centering
    \includegraphics[width=0.7\linewidth]{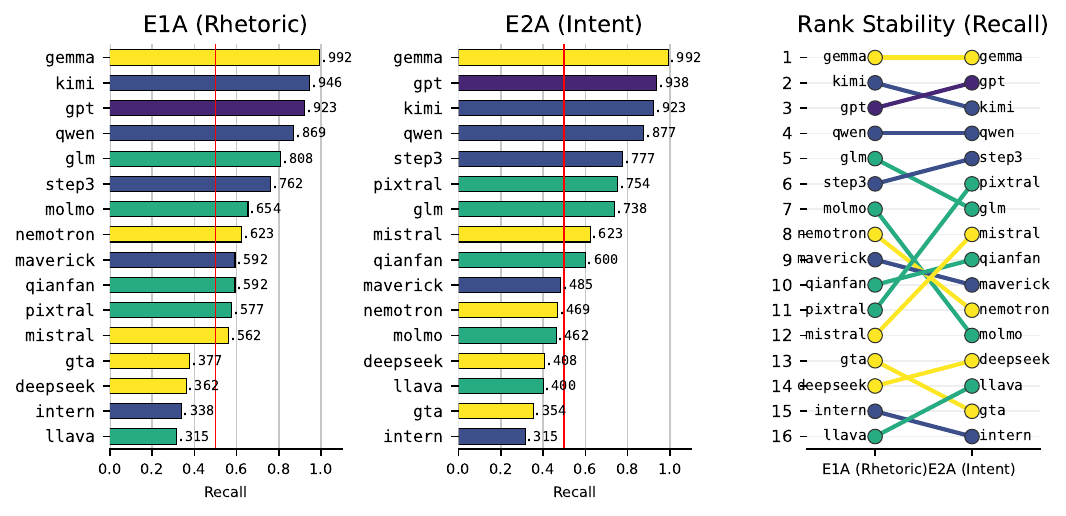}
    \caption{Per-model recall on VisLies (\experiment{1A} and \experiment{2A}). 
    High recall for \model{gemma} and \model{kimi} reflects positive-prediction bias rather than genuine detection competence; \model{llava}, \model{intern}, \model{deepseek}, and \model{gta} fail to flag a substantial share of unambiguous cases.}
    \label{fig:vislies_recall}
\end{figure*}

\begin{figure*}[h]
    \centering
    \begin{subfigure}{0.25\linewidth}
        \centering
        \includegraphics[width=\linewidth]{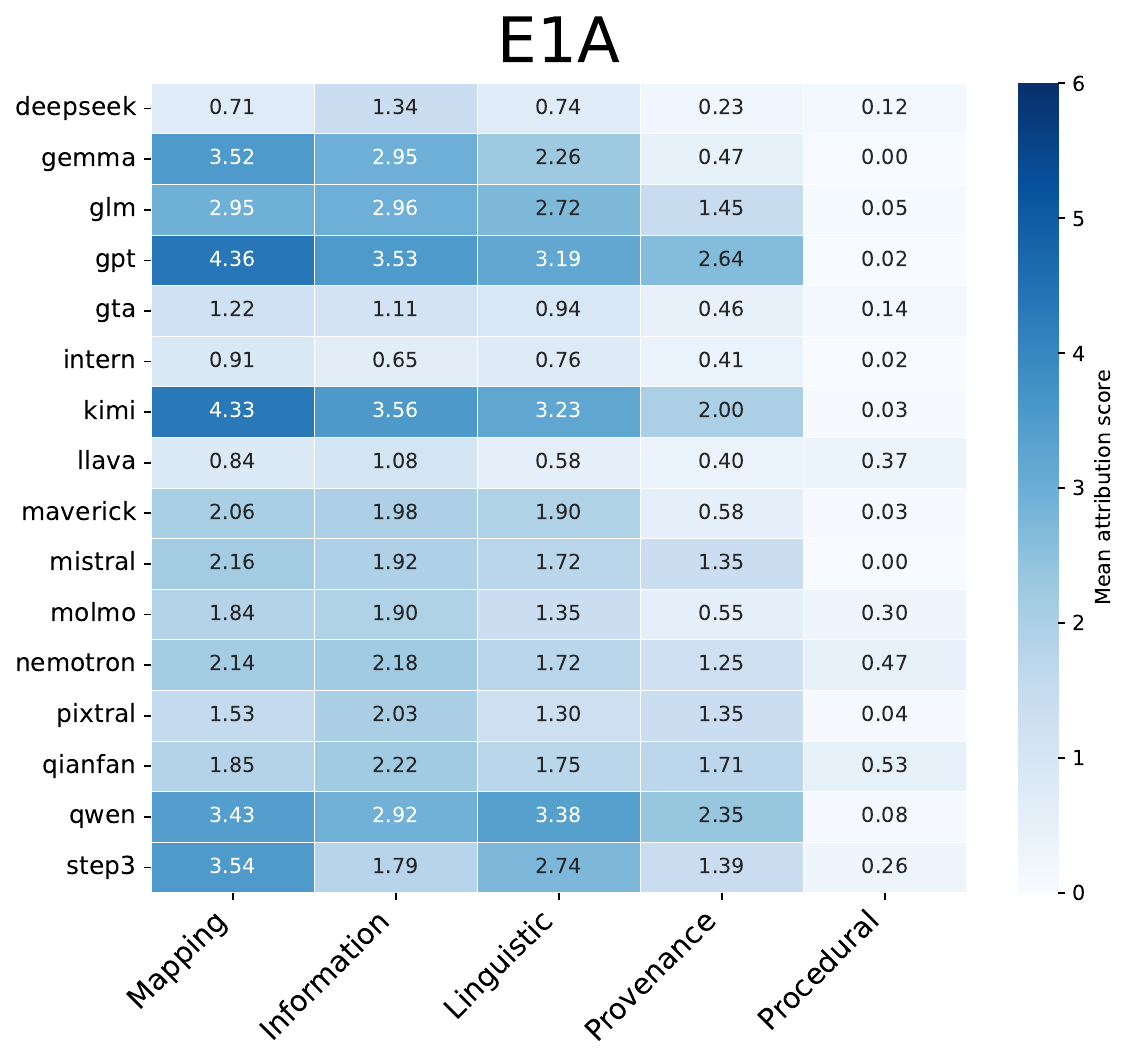}
        \caption{\experiment{1A}}
    \end{subfigure}
    \hfill
    \begin{subfigure}{0.25\linewidth}
        \centering
        \includegraphics[width=\linewidth]{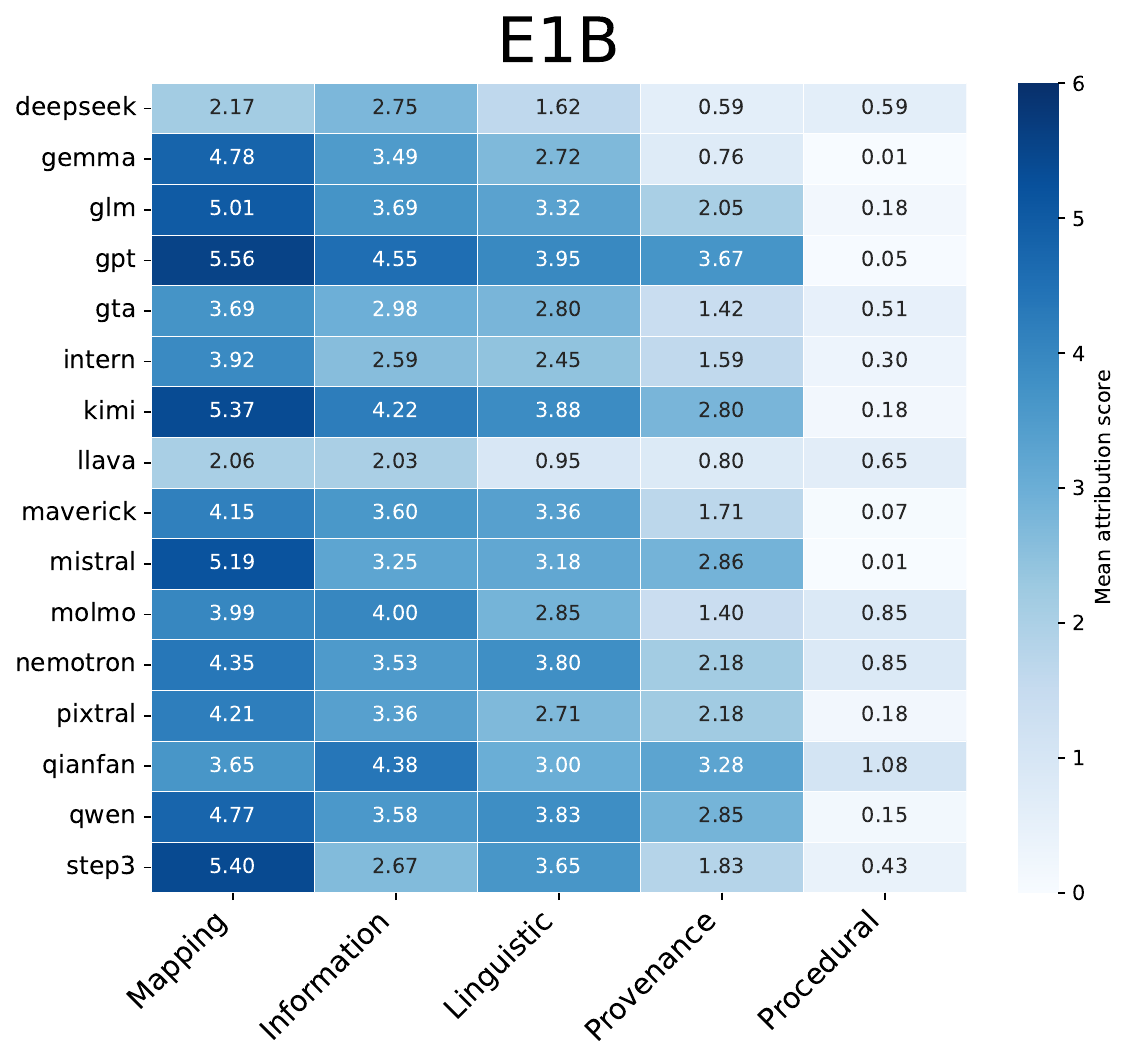}
        \caption{\experiment{1B}}
    \end{subfigure}
    \hfill
    \begin{subfigure}{0.25\linewidth}
        \centering
        \includegraphics[width=\linewidth]{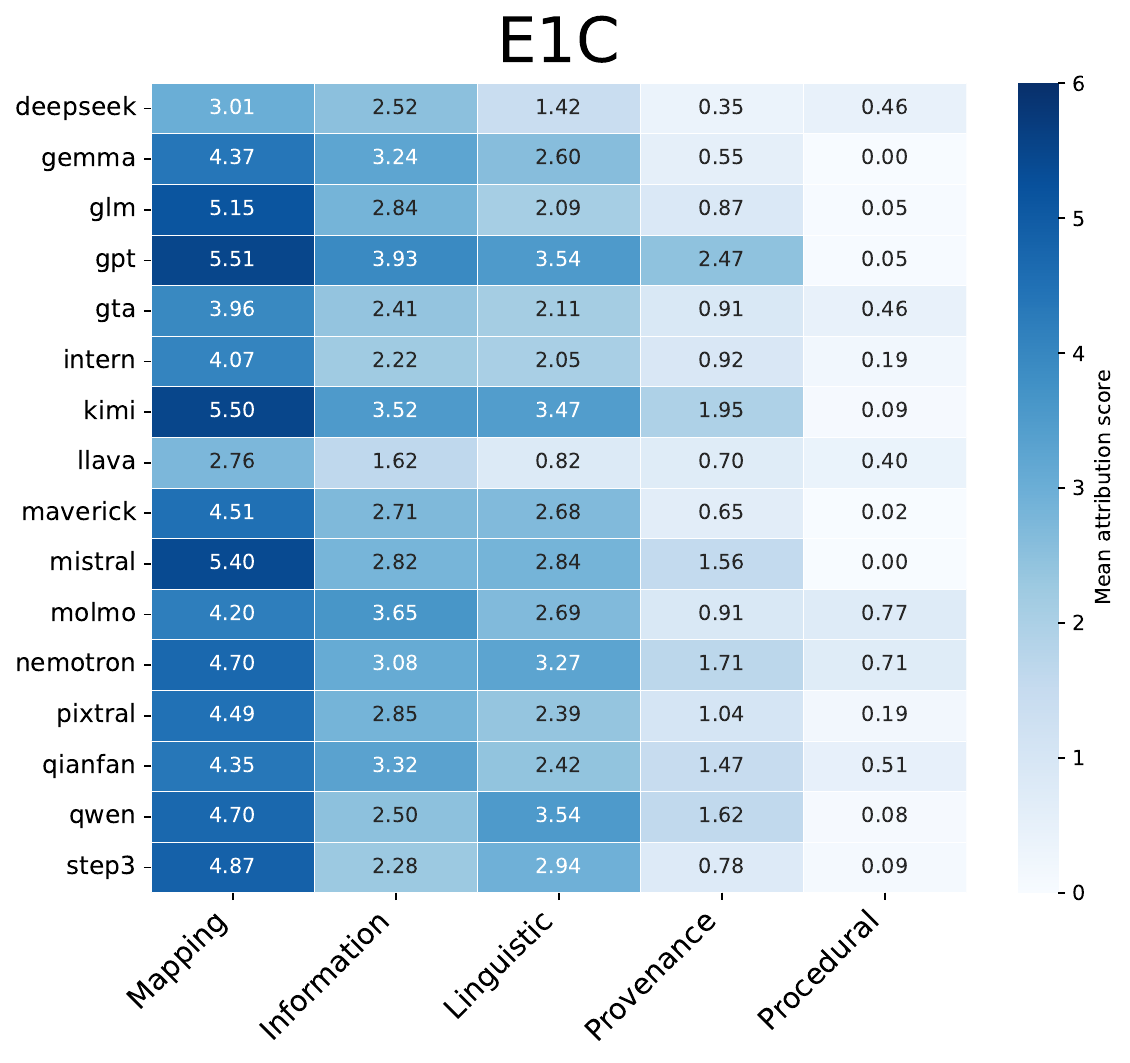}
        \caption{\experiment{1C}}
    \end{subfigure}
    \caption{Rhetoric attribution profiles on VisLies across conditions. \textit{Mapping} is the dominant rhetoric type in all conditions and for all models, in contrast to the COVID results, where \textit{Information Access} dominated. This inversion reflects the composition of VisLies, which is heavily populated by design-level distortions.}
    \label{fig:rhet_profile_vislies}
\end{figure*}

\begin{figure*}[h]
    \centering
    \begin{subfigure}{0.25\linewidth}
        \centering
        \includegraphics[width=\linewidth]{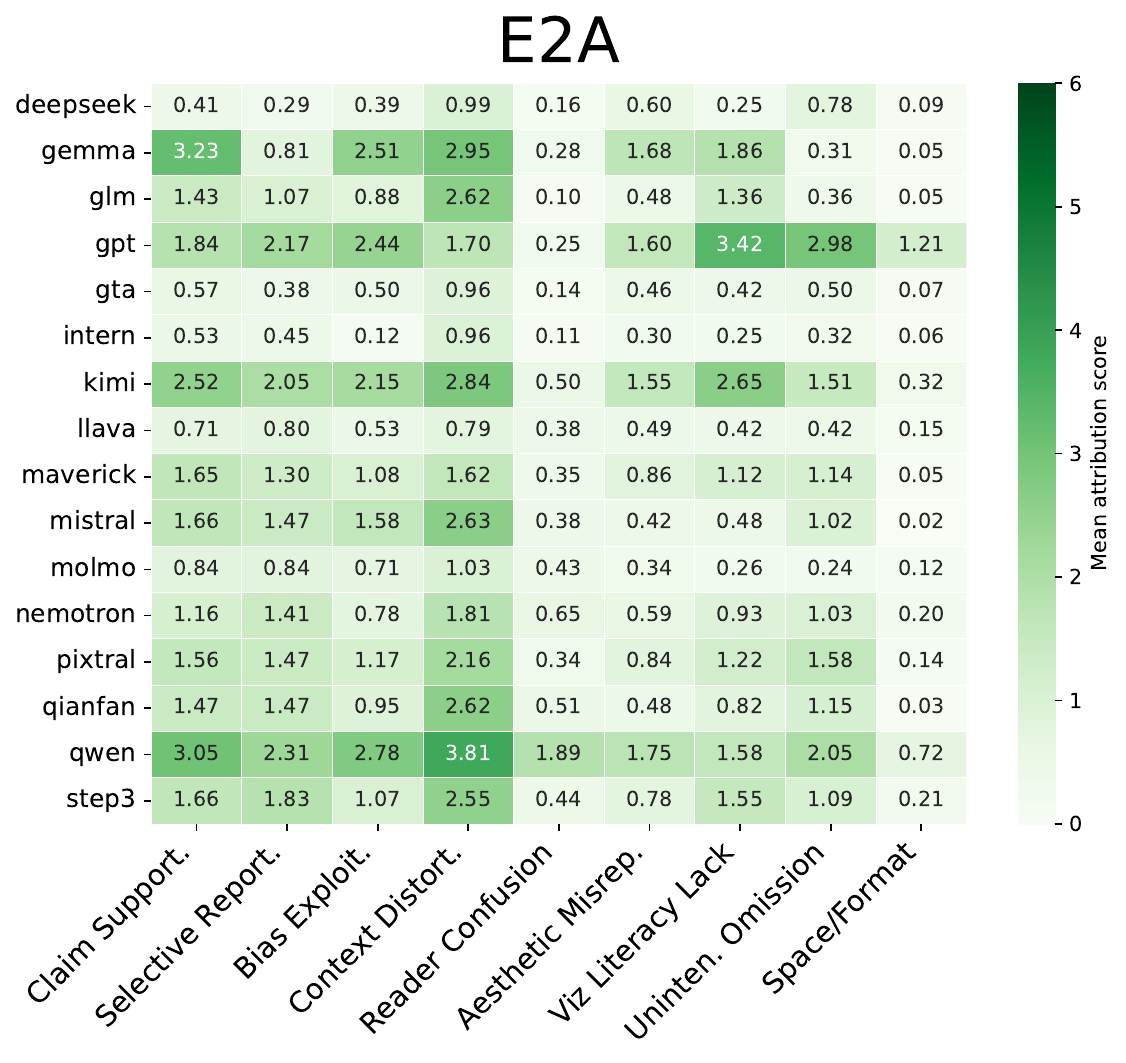}
        \caption{\experiment{2A}}
    \end{subfigure}
    \hfill
    \begin{subfigure}{0.25\linewidth}
        \centering
        \includegraphics[width=\linewidth]{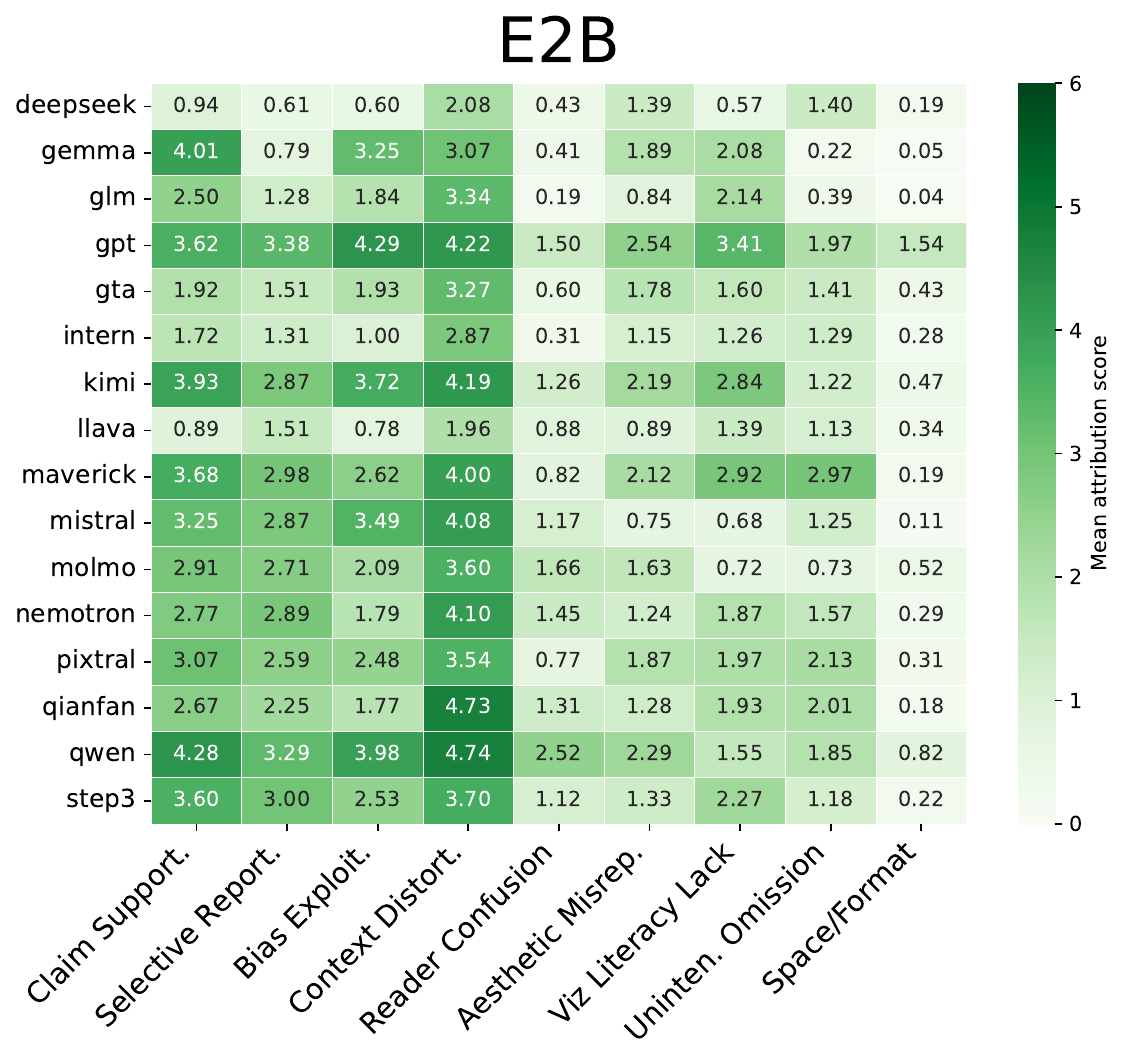}
        \caption{\experiment{2B}}
    \end{subfigure}
    \hfill
    \begin{subfigure}{0.25\linewidth}
        \centering
        \includegraphics[width=\linewidth]{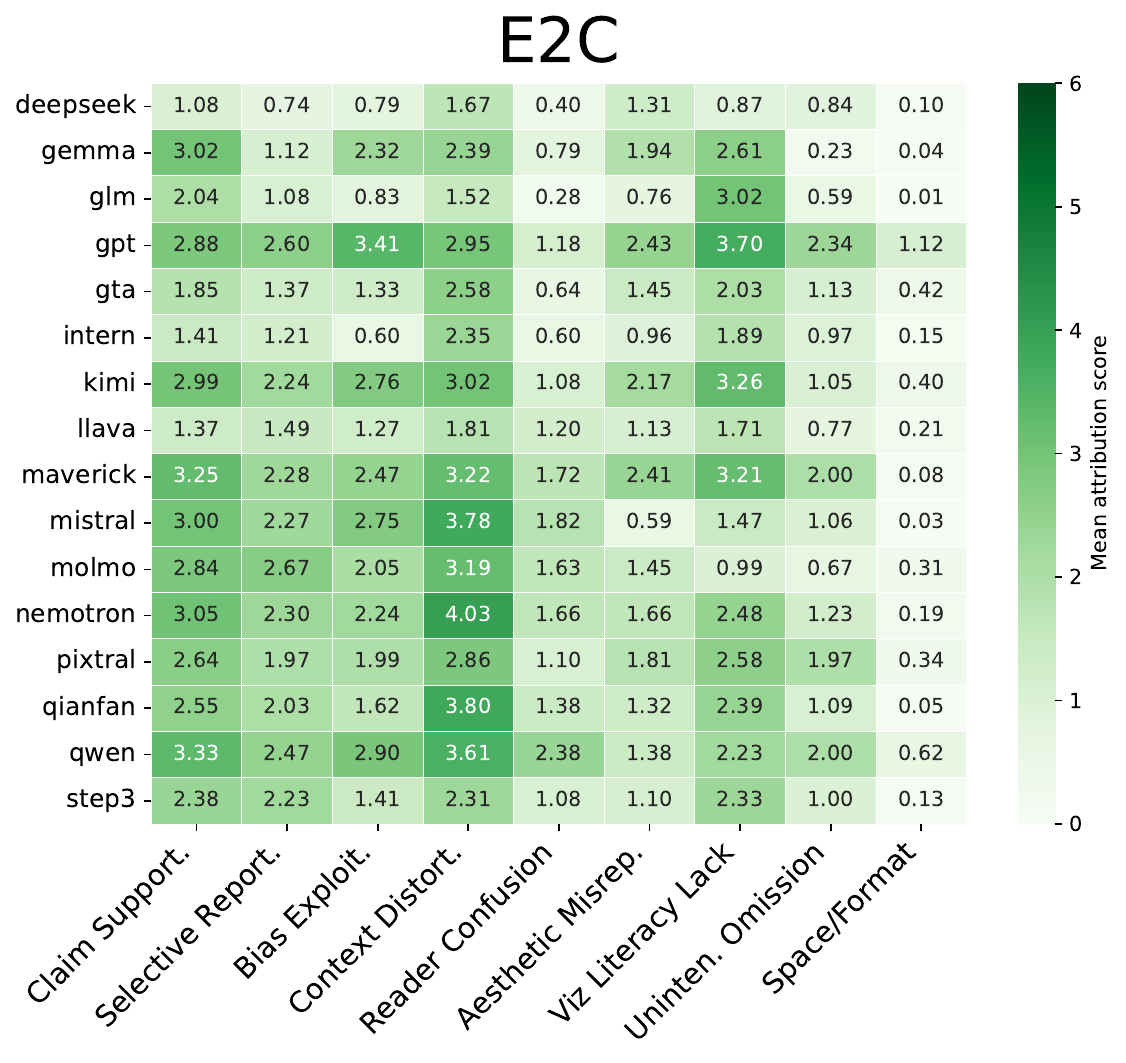}
        \caption{\experiment{2C}}
    \end{subfigure}
    \caption{Intent attribution profiles on VisLies across conditions. \textit{Context Distortion} is the dominant category, consistent across models and conditions.}
    \label{fig:intent_profile_vislies}
\end{figure*}

\end{document}